\documentclass{llncs}

\usepackage{amssymb}
\usepackage{amsmath}
\usepackage{mathtools}
\usepackage[sectionbib,sort]{natbib}
\usepackage{eurosym}

\usepackage[utf8]{inputenc} % allow utf-8 input
\usepackage[T1]{fontenc}    % use 8-bit T1 fonts
\usepackage{hyperref}       % hyperlinks
\usepackage{url}            % simple URL typesetting
\usepackage{booktabs}       % professional-quality tables
\usepackage{amsfonts}       % blackboard math symbols
\usepackage{nicefrac}       % compact symbols for 1/2, etc.
\usepackage{microtype}      % microtypography
\usepackage{cleveref}
\usepackage[font=small,skip=10pt,labelfont=bf,tableposition=top]{caption}
\usepackage[font=small,skip=0.3pt]{subcaption}
\usepackage[section]{placeins}

\newcount\Comments  % 0 suppresses notes to selves in text
\usepackage{color}
\definecolor{darkgreen}{rgb}{0,0.5,0}
\definecolor{purple}{rgb}{1,0,1}
\newcommand{\kibitz}[2]{\ifnum\Comments=1\textcolor{#1}{#2}\fi}
\newcommand*{\SuperScriptSameStyle}[1]{%
  \ensuremath{%
    \mathchoice
      {{}^{\displaystyle #1}}%
      {{}^{\textstyle #1}}%
      {{}^{\scriptstyle #1}}%
      {{}^{\scriptscriptstyle #1}}%
  }%
}

\usepackage{booktabs}

\newcommand*{\threeS}{\SuperScriptSameStyle{*{*}*}}

\usepackage{multirow}
\makeatletter
\newcommand\thefontsize{The current font size is: \f@size pt}
\makeatother

\usepackage{setspace}

\DeclareCaptionLabelFormat{andtable}{#1~#2  \&  \tablename~\thetable}

\title{Reinforcement Learning Policy Recommendation for Interbank Network Stability\thanks{We are grateful to the three anonymous reviewers and the editor, Dr. Iftekhar Hasan. Through their comments and suggestions the paper improved a lot during the revision process. However, any remaining errors or shortcomings are entirely our own.}}

\author{Alessio Brini\inst{1}\footnote{Corresponding author} \and Gabriele Tedeschi\inst{2,3} \and
	Daniele Tantari\inst{4}}
\authorrunning{Alessio Brini et al.} % abbreviated author list (for running head)
\tocauthor{Alessio Brini, Gabriele Tedeschi, Daniele Tantari}
\institute{Duke university, 305 Teer Engineering Building Box 90271, Durham, NC, 27708, United States, \\ 
	\email{alessio.brini@duke.edu}
	\and
	Università degli studi di Bari Aldo Moro - Piazza Umberto I - 70121 Bari ,Italy,
	\and
	LEE, Universitat Jaume I-Castellon, Spain
	\\ \email{gabriele.tedeschi@gmail.com}
	\and
		Università di Bologna - Via Zamboni, 33 - 40126 Bologna, Italy, \\
	\email{daniele.tantari@unibo.it}}

\begin{document}
\maketitle
\onehalfspacing
% \doublespacing
\section*{Abstract}
In this paper, we analyze the effect of a policy recommendation on the performance of an artificial interbank market. Financial institutions stipulate lending agreements following a public recommendation and their individual information. The former is modeled by a reinforcement learning optimal policy that maximizes the system's fitness and gathers information on the economic environment. The policy recommendation directs economic actors to create credit relationships through the optimal choice between a low interest rate or a high liquidity supply. The latter, based on the agents' balance sheet, allows determining the liquidity supply and interest rate that the banks optimally offer their clients within the market. Thanks to the combination between the public and the private signal, financial institutions create or cut their credit connections over time via a preferential attachment evolving procedure able to generate a dynamic network. Our results show that the emergence of a core-periphery interbank network, combined with a certain level of homogeneity in the size of lenders and borrowers, is essential to ensure the system's resilience. Moreover, the optimal policy recommendation obtained through reinforcement learning is crucial in mitigating systemic risk.

\newpage

\section{Introduction}\label{Sec:intro}
At the height of the sovereign debt crisis, the former president of the European Central Bank, Trichet, declared: ``When the crisis came, the serious limitations of existing economic and financial models immediately became apparent…As a policy-maker during the crisis, I found the available models of limited help. In fact, I would go further: in the face of the crisis, we felt abandoned by conventional tools…The key lesson I would draw from our experience is the danger of relying on a single tool, methodology, or paradigm. Policy-makers need input from various theoretical perspectives and a range of empirical approaches… In this context, I would very much welcome inspiration from other disciplines: physics, engineering, psychology, and biology. Bringing experts from these fields together with economists and central bankers is potentially very creative and valuable…'' (see \cite{BCE_T}).\\
Inspired by the words of Trichet, welcoming new and multidisciplinary policy tools, in this paper, we are explicitly interested in understanding the effect that an unconventional and environmentally dependent policy recommendation has on the stability of the interbank system. From the point of view of the functioning of the interbank market, our work follows \cite{berardi2017banks}, where financial institutions establish preferential lending arrangements to insure themselves against the unexpected withdrawal of deposits. Financial connections might change over time via a preferential attachment evolving procedure (see \cite{barabasi1999emergence}) such that each agent can enter into a lending relationship with others with a probability proportional to a fitness measure. Specifically, the attractiveness of agents is based either on their high supply of liquidity or their low interest rate. 
The authors show how implementing one or the other strategy generates different architectures of the credit network, which dissimilarly impact the spread of systemic risk.\\
The originality of this work with respect to the one mentioned above concerns the mechanism that drives banks to choose between the two strategies. Where in \cite{berardi2017banks} the choice is exogenous and fixed, here we introduce a time-dependent policy recommendation based on a reinforcement learning approach that directs banks to optimize the entire banking system's long-term fitness. Specifically, the regulator directs the interbank system towards an optimal strategy that chooses between favoring a high liquidity supply rather than a low interest rate, by collecting information from the environment.  %Then, depending on the regulator's recommendation, each bank signals in the market her optimal -microfounded- level of the indicated strategy used for the evolution of credit links.
Once the policy recommendation is made public, each bank signals to her counterparty within the interbank market her optimal liquidity supply or interest rate level, which are used to establish credit agreements via the above-mentioned preferential attachment mechanism. In a nutshell, we might think that the central bank directs the interbank system to choose between interest rate and liquidity supply by announcing the interest rate corridor that it publishes periodically. The corridor dynamics, therefore, influence the position of the Euro short-term rate (\texteuro STR) within it. Thus, indirectly, the position of the \texteuro STR in the corridor indicates the strategy the banks chose.
\\Compared to \cite{berardi2017banks}, therefore, the reinforcement learning mechanism allows us both to endogenize and identify the optimal strategy and to model a policy recommendation useful to tame systemic risk. Although this tool is helpful for modeling the reward-seeking behavior of agents in complex systems\footnote{ We refer the reader to \cite{charpentier2021reinforcement}, and \cite{mosavi2020comprehensive} for comprehensive reviews of different use cases of reinforcement learning in financial and economic contexts.} (see\citep{osoba2020policy}), to the best of our knowledge, it is barely employed in the agent-based framework. Interesting exceptions are \cite{liu2018agent}, and \cite{lozano2007artifitial}, which use reinforcement learning to model the credit allocation strategy of financial institutions in the interbank market. Apart from the modeling differences omitted here that distinguish us from those works, it is important to point out the methodological distinction. Where these works use a tabular reinforcement learning algorithm, as proposed by \citep{watkins1992q}, we use a state-of-the-art reinforcement learning algorithm with neural network approximators \citep{schulman2017proximal}, which describes the complex reward-seeking behavior. While the advantages and disadvantages of these algorithms are well documented and concern issues such as the computational efficiency, the curse of dimensionality, and the convergence \citep{bellman1956dynamic}, the better performance of the neural network-powered algorithms emerges. These models are beneficial when solving complex problems where the underlying environment changes rapidly and is also defined by the different forces that relate and compete with each other. These capabilities have already effectively solved complex financial and economic problems (see \citep{jiang2017deep,zhang2020deep,du2020deep,lin2020end}).\\ Without delving into technical details, some clarifications on how the proposed algorithm works should be done. The selected reinforcement learning algorithm optimizes an objective function that, in our context, corresponds to the aggregate fitness of the interbank system. The optimization is carried out by training a neural network model. The neural network receives input variables concerning the economic conditions of the interbank system and returns as output the strategy, i.e., the policy recommendation directing the system towards competing on liquidity supply rather than on the interest rate.

This family of algorithms is often criticized regarding the interpretability of inputs' impact on the results. The output, in fact, often appears as a black box whose determinants remain hidden from the user. To avoid this problem, we act in the following way. Firstly, we limit the choice of inputs to variables readily available to the regulator. To this end, we use aggregate systemic variables such as the interbank system's minimum, maximum, and average interest rate and liquidity supply. Choosing a limited set of input variables allows us to understand their effects in determining the output and to model a system with incomplete and asymmetric information (see \cite{bernanke1999financial}). Secondly, we directly study each input's impact on the output's determination through the SHapley Additive exPlanation (SHAP) framework \citep{lundberg2017unified}.

The introduction of the reinforcement learning framework into the interbank market model proposed by \cite{berardi2017banks} allows us to draw some important conclusions about the systemic stability of the system and to determine some policy interventions capable of curbing contagion. Firstly, the proposed algorithm fully endogenizes the evolution of the interbank network, whose architecture, therefore, changes over time. In this way, we can identify that the topology that emerges when the policy recommendation suggests a high supply of liquidity is more resilient in the face of exogenous shocks (see \cite{gai2010contagion} and \cite{elliott2014financial}, for similar results). Also, at the individual level, this policy produces better microeconomic performance. In this circumstance, the lenders and borrowers are more balanced in size, which generates a uniform risk exposure among counterparties able to favor the system's resiliency.  Although not unequivocally accepted (see, for instance, \cite{haldane2011systemic}), the negative impact of ''heterogeneity'' on systemic stability is in line with other theoretical and empirical studies (see \cite{caccioli2012heterogeneity}, \cite{iori2006systemic} and \cite{tedeschi2012bankruptcy}). On the other hand, the worse performance of a system dominated by low interest rates reflects the empirical evidence. Indeed, it is well documented that a credit market dominated by ''low-for-long'' interest rates adversely affects both the banks and the economy's stability. For financial institutions, low rates might reduce resilience by lowering profitability and, thus, their ability to replenish capital after a negative shock. This strategy would encourage risk-taking for the system, undermining systemic stability (see \cite{bindseil2018financial}, for a general overview of the topic).   
Finally, our results suggest that the policy recommendation implemented via reinforcement learning can more mitigate systemic risk than alternative tools.
\subsection*{Related literature}
The increasingly recurrent and impactful socio-economic crises have called for a deep rethinking of economic theory. Firstly, the literature has tried to understand and include the sources of contagion in the economic models. Regardless of the modeling approach used, which ranges from New Keynesian  models solved globally or using a reduced functional form (see, for instance, \cite{boissay2016booms}, \cite{gertler2020macroeconomic}, \cite{svensson2017cost}) to agent-based models and the most recent network-oriented approaches (see \cite{battiston2012default,battiston2012liaisons}, \cite{georg2013effect},\cite{haldane2011systemic}, \cite{upper2011simulation}, \cite{capponi2020dynamic}, \cite{calice2020contingent}), there is a general agreement that identifies interaction and heterogeneity as the drivers of endogenous crises. Moreover, the post-Lehman studies have placed particular emphasis on the propagation of contagion, determining the direction of the attack from financial to real markets and its fuse in the portfolio structure of financial institutions (see \cite{brunnermeier2012macroeconomics}). Many interesting studies, for example, have identified the source of contagion in the asset or liability side of banks' balance sheets. Among them, the effect of the fire-sale price and the (re)payment system between creditors and debtors have proven to be particularly important in generating financial instability (see \cite{acharya2008cash}, \cite{angelini1996systemic}, \cite{dasgupta2004financial} \cite{rochet1996interbank}).
In this vein, maturity transformation, sharing risk, herding behavior, and interbank linkages are just some of the various components able to trigger instability or collapse in financial markets (see \cite{acharya2008information},\cite{allen2000financial} and \cite{tedeschi2021macroeconomic}, among the many).
\\Once the origin of the disease and the channels through which it spreads have been identified, the literature has turned to treatment, that is, identifying the best tools to mitigate financial contagion. The scientific community has focused on developing new tools to overcome systemic instability. Several conventional and non-conventional monetary policies and other alternative tools have been proposed in this regard. However, their effects on financial stability are controversial and depend on the overall economic condition (see \cite{goldberg2020monetary}, and \cite{ECB2021}). A strand of literature, for example, has emphasized the importance of a strict, rule-based, and predictable monetary policy to tame systemic risk (see \cite{jimenez2014hazardous} and \cite{taylor2011macroeconomic}). On another side, instead, different studies have bet on alternative rules compatible with the underlying economic conditions (see \cite{boissay2021monetary}, \cite{grauwe2011banking} and \cite{gali2015monetary}). Unfortunately, the weak empirical evidence, due to the fairly recent development of these alternative techniques, which also include the so-called macro-prudential policies, makes it difficult to prove the supremacy of one approach over the other. While the empirical facts are still uncertain, recent theoretical models have attempted to resolve this ''certamen''. The model of \cite{boissay2021monetary} is an interesting contribution in this direction. The authors use a globally solved New Keynesian model with heterogeneous agents to generate endogenous crises. The paper compares two monetary policy instruments, one that follows a strict inflation-targeting rule and the other that allows the central bank to curb financial booms and busts. The authors show how the policies that mitigate output fluctuations help prevent financial crises by acting on agents' expectations. In support of cyclical policies determined by the economic background, there are also many agent-based models (see, \cite{cincotti2012macroprudential}, \cite{GIRI201942} and \cite{riccetti_russo_gallegati_2018}, among the many). Generating complex dynamics in evolving systems is an ideal environment for testing the effect of (un)conventional policies/measures on financial stability.
\newline

The rest of the work is organized as follows. In Section \ref{Sec:model} we present the functioning of the interbank market,  placing particular emphasis on the evolution of the credit network and the implementation of the reinforcement learning algorithm. In Section \ref{Sec:simulation}, we show the results. Specifically, we follow three steps: firstly, we verify the performance and robustness of the reinforcement learning algorithm; secondly, we investigate its implication on the interbank network morphology and the performances of the financial institutions; thirdly, we present the effect on the interbank systemic stability of the policy recommendation. Finally, Section \ref{Sec:fine} concludes with some remarks on the achieved results and the provided contribution.

\section{Model}\label{Sec:model}

This section describes the formation and evolution of credit relationships between financial institutions. Due to unexpected future movements of deposits,  banks enter into preferential lending agreements to have a potential credit channel when needed. These lending agreements are fast lanes created before use, and their set defines a {\it potential interbank network}. Banks report their credit conditions to their customers through an attractiveness measure to build their preferential lending agreements. We model bank fitness as combining a policy recommendation and private information. The first ingredient is a signal obtained via a reinforcement learning mechanism, through which the regulator directs banks to choose the best strategy given the underlying environmental conditions. In particular, the regulator recommends the weight to assign to high liquidity supply rather than to low interest rates, thus directing the competition. The second ingredient is a private signal, based on the bank's capital structure, consisting of the actual interest rate and credit provision offered. Potential credit relationships might change over time via a preferential attachment evolving procedure that depends on bank fitness. As the deposit shock materializes, financial institutions face liquidity surpluses or shortages, which induce them to exploit their preferential lending agreements and enter the interbank market as lenders or borrowers. At this point, the previously potential network becomes an  {\it active credit network}. Only the potential links of the banks facing a liquidity shortage are activated and correspond to a very sparse network. 
\subsection{The interbank market microstructure}\label{Subsec:microstructure}
We consider a sequential economy operating in discrete time, which is denoted by $t= \{0, 1, 2,\ldots,T\}$. At any time t, the system is populated by a large number $N$ of active banks $i,j \in \Omega=\{1,\ldots, N\}$. Financial institutions interact with each other through credit relationships represented by the set $V_t$, whose elements are ordered pairs of different banks. Banks (nodes or vertices) and their connections (edges or links) form the interbank network $G_t = (\Omega, V_t)$. The daily balance sheet structure of each bank is defined as 
\begin{equation}\label{Eq:balancesheet}
L_t^i+C_t^i+R_t^i=D_t^i+E_t^i,
\end{equation}
where assets are on the left-hand side and liabilities are on the right-hand one. In particular, $L, C$, and $R$ represent long-term assets, liquidity, and reserves, while $D$ and $E$ deposits and equity of bank i at time t. Reserves are a portion of deposits, $R_t^i=\hat{r}D_t^i$, where $\hat{r}$ is the required reserve rate\footnote{ This rate replicates a central bank regulation
that sets the minimum amount that a commercial bank must hold in liquid assets and is
commonly referred to as the reserve ratio. The central bank determines this minimum amount based on a specified proportion of bank deposit liabilities.}
% Specifically, deposits evolve into as follows: $D_t^i=D_{t-1}^i (1+\zeta \nu)$, with $\zeta$ to be a constant and $\nu$ a Gaussian white noise with zero mean and variance 1.

At every time $t$, deposits are exogenously shocked, and the balance sheet in Eq. \ref{Eq:balancesheet} modifies accordingly. Specifically, deposits evolve as 
\begin{equation}
   D_t^i=D_{t-1}^i (\mu+ \omega U(0,1)), 
\end{equation}
with  $U(0,1)$ a uniformly distributed noise between 0 and 1 and $\mu$ and $\omega$ modeling the expected number of negative shocks and thus different market conditions. On the one hand, financial institutions with a negative change in deposits and subject to a complete erosion of their liquidity become potential debtors in the interbank market. On the other hand, banks that suffer a small negative shock or an increase in deposits become potential creditors to the system\footnote{ It is worth pointing out that a positive (negative) deposit shock implies an increase (decrease) in reserves $R$. Consequently, banks plunder (replenish) their liquidity.}. Consequently, the respective demand $d^i_t$ and supply $s^i_t$ of liquidity of potential borrowers and lenders are given by

\begin{align*}
& \text{borrower if:} \Delta D_t^i+C_t^i \leq 0, \text{with demand of liquidity} d_t^i=\lvert \Delta D_t^i+C_t^i\rvert \nonumber \\
& \text{lender if}: \Delta D_t^i+C_t^i>0, \text{with supply of liquidity} s_t^i= \Delta D_t^i+C_t^i.
\end{align*}

Since we do not assume a Walrasian t\^atonnement mechanism, the system may endogenously generate a mismatch between credit supply and demand. Moreover, since the interbank network is not fully connected, even at a micro level, the borrower bank's liquidity demand might not match the credit supply offered by the lender banks connected to it. Specifically, we define the granted loan from a generic lender $i$ to a generic borrower $j$ as $l_t^{i,j}$ = $\min (s_t^i, d_t^j)$. Borrowing banks rationed in the interbank market can sell their long-term assets at a fire-sale price as a method of last resort. The amount of loan the borrower has to sell to cover its residual liquidity need equals $\Delta L_t^j = \frac{d_t^j-s_t^i}{\rho}$, where $\rho$ is the 'fire-sale' price. For the sake of simplicity and interoperability, modeling the stock market is out of the scope of our analysis. Therefore, we assume all other banks with sufficient liquidity to buy the same percentage of long-term assets from the distressed bank. This leads to an increase in the buyers' long-term assets and a decrease in their liquidity.

At the beginning of the next day, the repayment round takes place. Financial institutions encounter a new deposit movement that increases or decreases their liquidity. On the one hand, lending banks facing a positive (negative) change in deposits remain potential creditors (became potential debtors). On the other hand, borrowing banks face different scenarios depending on whether the deposit shock is positive or negative. Specifically, in the case of a positive shock, it can happen that: i) the change in deposits is sufficient to repay the principal and the interest, or  ii) the deposit variation is insufficient to cope with the loan. In the first case, the debtor can quickly meet her obligations, but in the second case, she must sell a number of long-term assets sufficient to repay the creditor at a fire-sale price fully.
On the other hand, in the case of a negative shock, banks must sell their long-term assets to pay for previous interbank borrowings and meet the new liquidity needs. All institutions that do not raise enough liquidity to meet their obligations via the fire sale fail, thus creating a bad debt for the lender. The creditor's loss, $B_t^{i,j}$, equals the granted loan after liquidating the debtor assets. Hence the equity of the bank $i$ obeys the following law of motion:
\begin{equation}\label{Eq:equity_motion}
E_t^i=E_{t-1}^i+\sum_j l_{t-1}^{i,j} r_{t-1}^{i,j}-\sum_{j\subseteq\theta_t^i} B_t^{i,j}-(1-\rho) \hat{L}_t^j, 
\end{equation}
where the second term on the right-hand side is the repayment, at the agent-specific interest rate $r^{i,j}$, of the granted loan $l^{i,j}$, and the third term is the bad debt of the subset of the bank $i$ clients, $\theta_t^i$, unable to repay their debts because they go bankrupt and the last term represents fire sales. If the bank has not fulfilled the loan requirements (i.e., if she cannot repay the principal and interest in full), the lender no longer provides credit, forcing her to exit the market. Thus, the borrower exits the market when assets fall short of liabilities, that is $E_t^i<0$. The failed banks leave the market. The banks exiting in $t$ are replaced in $t+1$ by new entrants, which are, on average smaller than incumbents. So, entrants' size is drawn from a uniform distribution centered around the mode of the size distribution of incumbent banks (see \cite{bartelsman2005comparative}).
\subsection{Banks microfoundations: the dynamics of lending agreements and trading strategies}\label{Subsec:dynamics}
At the beginning of each day, agents meet in the interbank market to meet their liquidity needs and sign bilateral potential lending agreements representing the directed links $(i,j)\in V_t$. These agreements can be interpreted as credit lines, which are valid during t and can be used at the request of the borrower $j$ in case of the lender $i$ available liquidity. The set of all potential lending agreements reproduces the potential interbank network topology\footnote{The creation of these links predates the deposit shock, which is why they are potential. These credit lanes, common in interbank markets, can be interpreted as mutual 'promises' of help between financial institutions in case of liquidity needs.}.\\ Let us now explain in detail the mechanism that governs the formation/evolution of credit relationships between financial institutions. We assume banks are risk-neutral agents operating in a perfect competition environment to optimize their expected profit. The bank $i$ expected profit for a loan provided to $j$ is given by
\begin{equation}\label{Eq:profit}
\mathbb{E}[\Pi_t^{i,j}]=p_t^{j}(r_t^{i,j} c_t^{i,j})+(1-p_t^{j}) (\xi A_t^{j}-c_t^{i,j})+\phi A_t^{j}-\chi A_t^{i},
\end{equation}
where $p_t^{j}$ is the probability that the borrower does not fail, $r_t^{i,j}$ the interest rate asked by the lender $i$ to the borrower $j$, $c_t^{i,j}$ the maximum amount $i$ is willing to lend to $j$. Moreover, $\xi$ is the liquidation cost of assets, $A_t^{j}$, pledged as collateral, and $\phi$ and $\chi$ the screening costs of creating a credit link that decrease with the debtor dimension and increase with the creditor size (see \cite{dell2004information}, and \cite{maudos2004factors}, for empirical evidence). Specifically, Eq. \ref{Eq:profit} captures the lender's expected revenue if the borrower does or does not meet her obligations (the first and the second term on the right side, respectively), and the opportunity cost of the agreement (last two variables in Eq. \ref{Eq:profit}). Moreover, we apply a heuristic rule to model a proxy for the debtor's $j$ survival probability. Recalling that the borrower fails if her equity becomes negative, $E_t^j<0$, the probability of surviving is given by the closeness between $ j$'s equity and the highest net worth in the system, i.e. 
\begin{equation}\label{Eq:probsurviving}
p_t^{j}=\frac{E_t^j}{E_t^{\max}}.
\end{equation}
 The bank's probability of surviving is connected to the financial fragility and the competition in an evolving financial system. On the one hand, a financial institution leaves the system if her net worth is so low that an adverse shock makes it negative or if she suffers a loss so huge as to deplete all the net worth accumulated in the past (see \cite{greenwald1993financial}). On the other hand, in a dynamic and competitive financial system, banks that do not keep up with their competitors have a higher probability of failing (see \cite{altman1993corporate}; \cite{denis1995causes}; \cite{10.2307/117389} and \cite{LANG199245}, among many). In this scenario, therefore, our bankruptcy probability predicts that if a bank remains too small compared to the competitors, her probability of failure increases (see \cite{altman2008value} \cite{DIETSCH2004773} \cite{altman2007modelling} and \cite{doi:10.1080/09603107.2014.896979}, for empirical evidence). The Eq. \ref{Eq:probsurviving} can also be interpreted as a rule of thumb for determining the risk premium that lenders charge to a borrower\footnote{We acknowledged that the simulation results are robust even when implementing a survival probability where $E_t^{\text{max}}$ does not change over time. Specifically, using the denominator of Eq. \ref{Eq:probsurviving} the average maximum equity over all the timestep.}.
 Finally, the maximum amount that the lender $i$ is willing to lend to $j$, that is, the lending capacity, $c_t^{i,j}$, in Eq. \ref{Eq:profit} is defined as
 $$
 \begin{cases}
& c_t^{i,j}= (1-h_t ^{j}) A_t^{j}>0$, if $(i,j)\in V_t,\\ 
& c_t^{i,j}=0  \ \hbox{otherwise},
\end{cases}
$$
with $h_t ^{j}\in (0, h_t ^{max})$ to be the borrower haircut, defined as the $j$'s leverage, $\lambda_t ^{j}$,  with respect to the maximum one. Hence $h_t ^{j}=\frac{\lambda_t ^{j}}{\lambda_t ^{\max}}$, with $\lambda_t ^{j}=\frac{L_t^j}{E_t^j}$. By setting Eq. \ref{Eq:profit} equal to zero and rewriting it as a function of $r_t^{i,j}$, we get the interbank rate that guarantees zero expected profit:
 \begin{equation}\label{Eq:interest}
r_t^{i,j}=\frac{\chi A_t^{i}-\phi A_t^{j}-(1-p_t^{j}) (\xi A_t^{j}-c_t^{i,j})}{p_t^{j}c_t^{i,j}}. 
\end{equation}
In line with the assumption of asymmetric information and costly state verification (see \cite{bernanke1999financial}), the lender applies an interest rate that increases with her size\footnote{The relationship between screening costs and the interest rate has been widely explored in the economic literature and often associated with the imperfect information paradigm (see \cite{10.2307/3989880}; \cite{10.2307/1821362}; \cite{hoff1990introduction})}. Following this interpretation, the explanation for the high interest rate lies in the problem of asymmetric information. Specifically, lenders having less information than borrowers about the latter's ability and willingness to repay a loan have to screen applicants and charge the cost of this operation to borrowers. However, it is infrequent to find evidence about the costs associated with screening and, more generally, about the effect of imperfect information on the behavior of credit market participants. (that is, her assets) and the financial vulnerability of the borrower (that is j's leverage). This last implication derives from the budget identity (see Eq. \ref{Eq:balancesheet}) from which we can derive that $A_t^{j}=\frac{L_t^{j}}{\lambda_t^{j}} + D_t^{j}$, where $\lambda_t^{j}=\frac{L_t^{j}}{E_t^{j}}$. In addition, the interest rate in Eq. \ref{Eq:interest} is not linearly related to the bank's survival probability and capacity.

We now have all the elements to describe how traders select their counterparts in the interbank system, i.e., how lending arrangements are formed and evolve. We develop a measure of agent attractiveness to generate an endogenous preferential attachment mechanism. Specifically, banks signal themselves to their pool of clients based on their low interest rates or abundant supply of liquidity.  
%Specifically, banks try to signal themselves in the interbank market by offering low-interest rates or conspicuous liquidity supplies. 
The dichotomy between these two strategies is microfounded and stems from the expected profit of banks (see Eq. \ref{Eq:profit}), where the screening costs of creating a credit link increase with the creditor size. This implies, as shown in Eq. \ref{Eq:interest}, that lenders attractive in terms of higher liquidity supply offer higher interest rates. Symmetrically, banks offering low interest rates are necessarily less liquid\footnote{Assuming screening costs that increase with borrower's dimension and decrease with the lender's dimension implies an inverse relationship between the lender's size and the interest rate the
financial institution offers on the interbank market: the most liquid lender provides the best conditions in terms of interest rate. In this circumstance, the two banks' strategies collapse into the same. Since the banks' strategies go in the same direction, their impact on the simulated dynamics is similar. Given the perfect overlap of the two tactics, the reinforcement learning mechanism achieves precisely the same effects as a random choice.}. The positive relationship between the liquidity of financial institutions and their interest rates also has empirical evidence. Indeed, in the presence of distortions in the functioning of the financial market due to increasing heterogeneity in the agents' size (see \cite{10.1093/rfs/hhr018} and \cite{RePEc:tiu:tiucen:62237388-9a7c-458c-8608-957e61e2adb1}) and the segmentation of the market itself (see \cite{veyrune2018relationship}), the modeled positive correlation between the two variables emerges.

Although all agents start from the same initial conditions, financial institutions are characterized by heterogeneous levels of their agent-specific variables as time goes by. In line with this, the fitness of each agent $\mu_t^i$ is a combination of her liquidity relative to the highest liquidity provided in the market, $C_t^{\max}$,  and her interest rate compared to the cheapest one, $r_t^{min}$, i.e.  
\begin{equation}\label{Eq:fitness}
\mu_t^i=\eta_t\biggl(\frac{C_t^{i}}{C_t^{\max}}\biggr)+(1-\eta_t)\biggl(\frac{r_t^{min}}{r_t^{i}}\biggr).
\end{equation}
The parameter $\eta_t$ reflects a policy recommendation at time t, addressing the choice of the banking sector towards one of two possible strategies. On the one hand, $\eta$ approaching zero identifies an interbank system moving towards the cheapest interest rates. On the other hand, $\eta$ close to one highlights a liquidity-based system. The signal disseminated by the regulator that directs the system toward the optimal strategy can be interpreted as the central bank's announcement of the interest rate corridor. This corridor conditions the interbank interest rate and, consequently, the choice of each financial institution on her credit condition (see \cite{giannone_lenza_pill_reichlin_2011}).  We refer the reader to Subsection \ref{Subsec:learning} for a detailed explanation of the policy recommendation evolution. One of the main contributions of our work is to assume  $\eta_t$  endogenously evolving through a reinforcement learning mechanism, modeling the regulator's will to address the banking system toward the best credit strategy for system stability. It is worth emphasizing that, although in Eq.\ref{Eq:fitness} the public signal is homogeneous in the baseline model, banks' attractiveness remains highly heterogeneous as the private signals on the liquidity, $C^i_{t}$, and interest rate, $r^i_{t}$, are agent-specific. Let us assume, for example, that the system is directed towards a low-interest rate, $\eta=0$. Since interest rates in the fitness measure are bank-specific, interest rates applied by lenders to their clients are different. Furthermore, the liquidity supply of those lenders chosen to grant credit is also agent-specific, which ensures heterogeneity in granting credit. A similar dynamic applies to the case where the signal directs toward a high liquidity supply, i.e., $\eta=1$. In other words, the only element of homogeneity is the public signal that directs the system toward the optimally selected strategy\footnote{This assumption is modified in Sec \ref{herd}, where heterogeneity is also introduced in the public signal.}

Regarding our interbank network model, credit links are directional because they are created and deleted by the agent $j$, who looks for a loan and points to the agent $i$ that provides credit. %\footnote{Banks are divided between potential lenders and borrowers according to the sign of the deposit shock described in Sec. \ref{Subsec:microstructure}}. 
The information on credit conditions (and then loan) flows the opposite.  It is worth noting that credit terms are bilateral (between creditor and debtor) and, therefore, not available from other market members.\\ In general local interaction models, the agent interacts directly with a finite number of counter-parties in the population. The set of nodes with which a single node is linked is called its neighborhoods. In our model, the number of outgoing links is constrained to be a small number $\hat{d}$. Thus borrowers can only get loans from $\hat{d}$ lenders. With this assumption of network sparsity, the topology is always locally tree-like, avoiding loops that would preclude us from fully understanding the network architecture's impact on economic dynamics, such as systemic risk, failures, and liquidity diffusion.\\ At the time $t=0$, each bank $j$ starts having $\hat{d}$ random outgoing links (i.e., potential borrowing positions) and possibly with some incoming links from other agents (i.e., potential lending position). At the beginning of each period, links are rewired in the following way. For any outgoing link $i$, each borrower $j$ randomly selects a new bank $k$. Comparing the  fitness of the new financial institution with the one of its previous lender $i$, the borrower $j$ cuts her old link with $i$ and creates a new one with $k$ according to the probability 
\begin{equation}\label{Eq:probattachment}
P_t^j=\frac{1}{1+e^{-\beta(\mu_t^k-\mu_t^i)}},
\end{equation}
or keep its previous link with probability $1-P_t^j$. The proposed mechanism for reviewing credit agreements ensures that the most attractive lenders get the highest number of borrowers (i.e., incoming links) and earn the highest profits. Nevertheless, the degree of randomness in the algorithm guarantees that some links with very high-performing agents may be cut in favor of less attractive creditors. The amount of randomness is regulated by $\beta$ and has a double purpose: from a practical point of view, it prevents the system from being centralized around a single financial hub; from a theoretical perspective, it allows us to model incomplete information and bounded rationality.

%%%%%%%%%%%%%%%%%%%%%%%%%%%%%%%%%%%%%%%%%%%%%%%%%%%%
\subsection*{The evolution of the banking system: determining the policy recommendation}\label{Subsec:rl}
As anticipated in the previous section, we use the reinforcement learning paradigm to move the parameter $\eta_{t}$ and obtain an optimal policy recommendation in the described banking system. Reinforcement learning aims to solve a decision-making problem in which the timing of costs and benefits is relevant. In an interbank market that follows the specified dynamics for the creation of lending agreements, reinforcement learning can help determine the policy recommendation that better identifies the optimal attachment strategy to follow in Eq. \ref{Eq:fitness}, even when partial information about the system is provided. Hereafter, we refer to the reinforcement learning algorithm as the learning algorithm. A Markov Decision Process (MDP) is the mathematical formalism under which the reinforcement learning problem is usually defined. A MDP comprises of a set of possible states $S_{t} \in \mathcal{S}$, a set of possible actions $A_{t} \in \mathcal{A}$ and a transition probability  $P[S_{t+1} = s' \mid S_{t} = s, A_{t} = a]$. At each time $t$, a learning agent that is in state $S_{t}$, takes an action $A_{t}$ and receives a reward $R_{t+1}\left(S_{t}, A_{t}, S_{t+1}\right) \in \mathbb{R}$ from the environment before moving to the next state $S_{t+1}$. We define the agent  strategy $\pi: \mathcal{S}\times \mathcal{A} \mapsto [0,1]$ as the conditional  probability $\pi(a \mid s)$ of taking the action $A_{t}=a$ being in the state $S_{t}=s$. The reinforcement learning problem is the stochastic control problem of maximizing the expected discounted cumulative reward
\begin{equation}\label{Eq:objective}
\mathbb{E}_{\pi}\left[\sum_{t=0}^{\infty} \gamma^{t} R_{t+1}\left(S_{t},A_{t},S_{t+1}\right)\right],
\end{equation}
where $\gamma \in [0,1)$ is a discount factor, and the expectation is w.r.t. the sequence of states and actions reached following the strategy $\pi$.

In our MDP, the sequential economy in which the banking system operates plays the role of the environment. Banks interact with the environment by changing their credit lines: each day, they can adapt their attachment strategy between liquidity supply and interest rate discount, which is regulated through Eq. \ref{Eq:fitness}, with the choice of $\eta_{t}$, playing the role of the action $A_t$.   We assume the agent is the system as a whole rather than the single bank and that the optimal strategy is realized at the system level, i.e., that the regulator directs financial institutions towards the correct combination of the two strategies. This assumption has a twofold purpose. On the one hand, it helps us to model a system with incomplete/asymmetric information, where the central bank has richer information set than the single economic actor (see, for instance, \cite{hoff1990introduction}, and \cite{thakor2020fintech}). On the other hand, it allows us to incorporate economic policy, seen as the optimal indication that the regulator gives to the system to reduce the interbank market vulnerability (see \cite{BCE_T}, for a global overview)
\footnote{ Considering $\eta$ as a system variable allows us to reduce the problem's mathematical and computational complexity and study the banking system's behavior as a whole. Making $\eta$ bank specific leads towards multi-agent reinforcement learning applications (\cite{bucsoniu2010multi}), which consider agents that compete with each other and are an out-of-the scope of the present paper.}. As shown above, the central bank's recommendation is made through the optimal interest rate corridor announcement, which conditions interest rates and the liquidity supplied by financial institutions.

The state $S_t$ includes information on the banking system's liquidity $C_{t}$ and the interest rate $r_{t}$ distributions. Specifically, the state space is defined as  $$S_{t}=(C_{t}^{\max},C_{t}^{\min},r_{t}^{\max},C_{t}^{\text{avg}},r_{t}^{\min},r_{t}^{\text{avg}}),$$ where $x_{t}^{\max} = \max_{i\in\Omega} x_{t}^{i}$, $x_{t}^{\min} = \min_{i\in\Omega} x_{t}^{i}$, $x_{t}^{\text{avg}} = \sum_{i=1}^{N} x_{t}^{i}/N$, being $x$ the variable of interest. We believe that this state-space setting is realistic enough to model the partial information of the regulator about the banking system:  it would be difficult and costly to retrieve detailed and specific data on all the banks included in the system at each time step. It is easier to gather information about the best and the worst liquidity provider in the interbank network as much as average estimates of the entire market.

Finally, the reward function we consider is the system's total fitness
\begin{equation}\label{Eq:reward}
    R_{t}\left(S_{t},A_{t},S_{t+1}\right) = \sum \limits_{i=1}^{N} \mu^{i}_{t}
\end{equation}
 Moreover, the problem in Eq. \ref{Eq:objective} becomes a maximization of the discounted cumulative banks' total fitness. From the definition of bank fitness, this means guaranteeing a better flow of liquidity through the banking system and an efficient allocation at a more convenient interest rate. We recall here that maximizing the fitness of financial institutions corresponds to optimizing their expected profit. The motivation behind this modeling assumption is twofold. Firstly, for the recommendation to be followed by the banks, it must have a goal of interest to the banks themselves, namely their profit. Second, the regulator, by maximizing the fitness of the system, succeeds ex-post in safeguarding the resilience of the financial system, given the inverse relationship between expected profits and failures of financial institutions.
 %Once found the optimal strategy, at each time step, the learning agent (the regulator) observes the state of the system $S_{t}$ and chooses the best action $A_{t} = \eta_{t}$ from  $\mathcal{A}$ in order to reach a system configuration which is both practical and efficient at the same time. 
 
 The learning algorithm operates in a model-free setting because it only receives partial information on the relevant variables of the system. At the same time, it does not know the internal dynamics (i.e., transition probability) with which the banks' balance sheets move and lending agreements are generated. This information has to be inferred through the sequence of states, actions, and rewards during the learning process.

\subsection{The optimization algorithm: Proximal Policy Optimization}\label{Subsec:ppo}

The optimization problem in Eq. \ref{Eq:objective} can be solved using a policy gradient algorithm like the Proximal Policy Optimization (PPO) \citep{schulman2017proximal}. A policy gradient algorithm directly parametrizes the optimal strategy  $\pi_{\theta} = \pi(a\mid s; \theta)$, for example, using a multilayer neural network with  parameters $\theta$. 
%The parametrization of the policy allows the agent to select an action without consulting a value function, unlike a value-based method. 
The optimization problem is approximately solved by computing the gradient of the cumulative fitness of the system  $J(\theta) = \sum_{t=0}^\infty \gamma^t R_{t+1}(S_t,A_t,S_{t+1};\pi_{\theta})$ and then carrying out gradient ascent updates according to 
\begin{equation}\label{Eq:PGupdate}
    \theta_{t+1}=\theta_t +\alpha \nabla_{\theta} J(\theta_{t}), 
\end{equation}
where $\alpha$ is a scalar learning rate. The policy gradient theorem (\cite{sutton2000policy,marbach2001simulation}) provides an analytical expression for the gradient of $J(\theta)$ as
\begin{align}\label{Eq:pgtheorem}
    \nabla_{\theta} J(\theta)  &= \mathbb{E}_{\pi_{\theta}} \left[ \frac{\nabla_{\theta} \pi\left(A_{t} \mid S_{t};\theta \right)}{\pi\left(A_{t} \mid S_{t};\theta \right)}  Q_{\pi_\theta}(S_{t}, A_{t}) \right]  \\ \nonumber
    &= \mathbb{E}_{\pi_{\theta}} \left[  \nabla_{\theta} \log\pi\left(A_{t} \mid S_{t};\theta \right)  Q_{\pi_\theta}(S_{t}, A_{t}) \right],
\end{align}
where the expectation, with respect to $(S_t,A_t)$, is taken along a trajectory (episode) that occurs adopting the strategy $\pi_{\theta}$ and the action-value function
\begin{equation}\label{Eq:Qfunc}
Q_{\pi}(s, a) \equiv \mathbb{E}\left[\sum_{k=0}^\infty  \rho^k R_{t+1+k} \mid S_{t}=s, A_{t}=a, \pi\right], 
\end{equation}
represents the long-term reward associated with the action $a$ taken in the state $s$ if the strategy $\pi$ is followed hereafter.
%Following this gradient means updating the strategy's parameters towards those actions that yield the highest sum of rewards. 
It can be proven that it is possible to modify the action value function $Q_\pi(s,a)$ in ($\ref{Eq:pgtheorem}$) by subtracting a baseline that reduces the variance of the empirical average along the episode while keeping the mean unchanged. A popular baseline choice is the state-value function
\begin{equation}\label{Eq:Vfunc}
    V_{\pi}(s) \equiv  \mathbb{E}\left[\sum_{k=0}^\infty  \rho^k R_{t+1+k} \mid S_{t}=s, \pi\right], 
\end{equation}
which reflects the long-term reward starting from the state $s$ if the strategy $\pi$ is adopted onwards. The gradient thus can be  rewritten as
\begin{equation}\label{Eq:pgadvantage}
    \nabla_{\theta} J(\theta)  = \mathbb{E}_{\pi_\theta} \left[  \nabla_{\theta} \log\pi\left(A_{t} \mid S_{t};\theta_{t} \right)  \mathbb{A}_{\pi\theta}(S_{t}, A_{t}) \right]
\end{equation}
where 
\begin{equation}\label{Eq:Afunc}
    \mathbb{A}_{\pi}(s, a) \equiv Q_{\pi}(s, a)-V_{\pi}(s),
\end{equation}
is called advantage function and can be interpreted as the gain obtained by choosing a specific value of $a$ in a  given state with respect to its average value for the strategy $\pi$.

Different policy gradient algorithms derive from the way the advantage function is estimated. In PPO, the advantage estimator  $\mathbb{A}\left(s,a;\psi\right)$ is parameterized  by another neural network with  parameters $\psi$. This approach is known as actor-critic: the actor is represented by the policy estimator $\pi(a|s;\theta)$ that outputs a probability for each possible value of $a\in \mathcal{A}$, which the learning algorithm uses to sample actions, while the critic is the advantage function estimator $\mathbb{A}\left(s,a;\psi\right)$ whose output is a single scalar value. The two neural networks interact during the learning process: the critic drives the updates of the actor, which successively collects new sample sequences that will be used to update the critic and again evaluated by it for new updates. The extended objective function can therefore describe the PPO algorithm 
\begin{equation}\label{Eq:ppoobj}
    J^{\text{PPO}}(\theta,\psi)=  J(\theta)-c_{1} L^{\text{AF}}(\psi) +c_{2} H\left(\pi\left(a \mid s; \theta\right)\right)
\end{equation}
where the second term is a loss between the advantage function estimator $\mathbb{A}\left(s,a;\psi\right)$ and a target $\mathbb{A}^{targ}$, represented by the cumulative sum of discounted reward, needed to train the critic neural network. The last term represents an entropy bonus to guarantee an adequate level of exploration. Details about the specific choice of the target, together with additional information about the general algorithm implementation, are given in the App \ref{App:B}. In what follows, PPO can be generally referred to as the learning algorithm.

%%%%%%%%%%%%%%%%%%%%%%%%%%%%%%%%%%%%%%%%%%%%%%%%%%%%
\section{Simulation Results}\label{Sec:simulation}
In this section, we perform numerical experiments to test the capability of the learning algorithm to identify an optimal strategy for selecting $\eta$ and trading off the two competing ways of establishing credit relationships. In this respect, we analyze the effects of the $\eta$ dynamics on agents' economic performances, the interbank network topology, and its resilience in the face of exogenous shocks. Finally, we study the effect of the policy recommendation obtained through reinforcement learning in controlling credit crunch phenomena and mitigating systemic risk. 

The results provided in the following subsections are obtained from simulated tests, which share some choices for the parameter involved in the dynamic simulation of the system. The number of Monte Carlo simulations performed is $M=200$, and each simulation is $T=1000$ periods long. We simulate a system with $N=50$ banks whose out-degree is $\hat{d}=1$, so each bank can obtain at most one outgoing link at each time step while can have many possible incoming links. Each bank is subjected to an initial probability of being isolated, set at 0.25. The parameters of the screening cost $\chi$, and $\phi$ that enter in Eq. \ref{Eq:interest} are set respectively at 0.015 and 0.025, while the liquidation cost of collateral $\xi$ is 0.3. The parameters $\mu$ and $\omega$ shifting the uniformly distributed noise that shocks the bank deposits are set at 0.7 and 0.55. All the banks start with the same initial interest rate equal to $2\%$ and are endowed with the same initial balance sheet $C_0 = 30$, $L_0 = 120$, $D_0 = 135$, and $E_0 = 15$.  The reserve ratio $\hat{r}=0.2$, the price of fire sale $\rho=0.3$ and the intensity for breaking the connection between banks $\beta=5$ in Eq. \ref{Eq:probattachment} are other parameters common to all the agents in the network. In the App. \ref{App:A} we check the robustness of our qualitative results by changing some key parameters. Specifically, we vary the intensity of choice, $\beta$, from 0 to 40 with steps of 2; the fire-sale price, $\rho$, from 0.1 to 0.5 with steps of 0.1, the reserve rate, $\hat{r}$ from 1\% to 10\% with steps of 0.1 and, finally, the parameter $\omega$ regarding the volatility shock on bank deposit. We have then studied the moments of the distributions of the statistics of interest. Results confirm that our findings are robust to some variations of the banking system simulation.
% The random shock of the bank deposits is distributed as a shifted uniform distribution $\mu + \eta\mathbb{U}(0,1)$ where $\mu=0.7$ and $\eta=0.55$. This choice guarantees a random shock slightly skewed towards negative shocks to encourage the bank in the system to ask for liquidity and obtain a natural flow of money.

The PPO algorithm parametrizes a discrete strategy function so that the learning algorithm can choose the value of $\eta$ among a finite set of actions $\mathcal{A}=\{0,0.5,1\}$\footnote{Under the same setting, training PPO instances that are allowed to pick fine-grained discrete values between 0 and 1 as a possible action is computationally expensive because the algorithm needs to explore a broader set of possible state-action pairs. Such an implementation would let the algorithm runtime grow and would not necessarily improve the results because the algorithm would not be able to alias between consecutive actions. A fine-grained action space $\mathcal{A}$ would make the $\eta$-strategy less interpretable. Hence in our analysis, we decided to distinguish three specific scenarios, which are the two extreme cases ($\eta=0.0$ and $\eta=1.0$) and the middle case ($\eta=0.5$).}

%%%%%%%%%%%%%%%%%%%%%%%%%%%%%%%%%%%%%%%%%%%%%%%%%%%%%%%%%%%%%%%%%%%%%%%%%%%%%%%%%%%%%%%%%%%%%
\subsection{Training the PPO algorithm}\label{Subsec:learning}
As the first step in our numerical analysis, we evaluate the performance of the strategy learned by the PPO algorithm. We train four PPO instances on $E_{in}=1000$ consecutive episodes, which are independent simulations of the banking system. The PPO instances differ for the random seed used to initialize the neural networks and to train them using a stochastic gradient descent approach. Multiple concurrent training of different instances is needed to provide an average performance together with a confidence interval that highlights the robustness of the learning process. Each training episode consists of a simulation of the banking system for $T$ periods that allow the learning algorithm to collect samples of data with which it can perform updates of the model parameters. During the learning phase, we evaluate the learning progress of each instance at several intermediate steps. We fix the weights of the neural networks that parametrize the $\eta$ public signal and perform $E_{out}=5$ out-of-sample test episodes before carrying on the training process to assess the learned behavior up to that point. We refer to the App. \ref{App:B} for the technical difference between an in-sample and an out-of-sample test episode.

\begin{figure}
\centering
\begin{subfigure}{.5\textwidth}
 \centering
  \includegraphics[width=\textwidth]{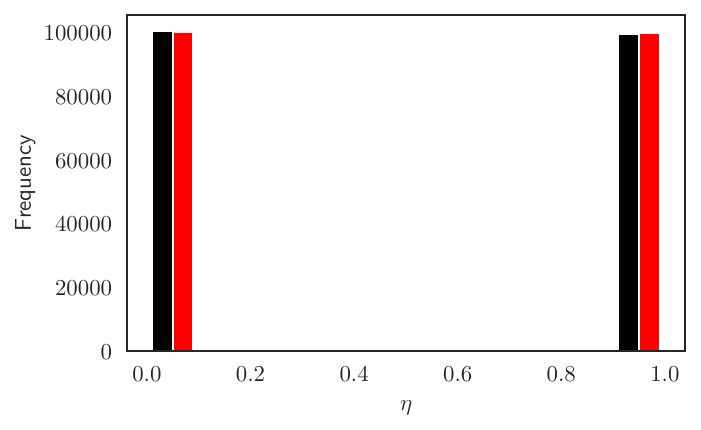}
\end{subfigure}%
\begin{subfigure}{.5\textwidth}
 \centering
  \includegraphics[width=\textwidth]{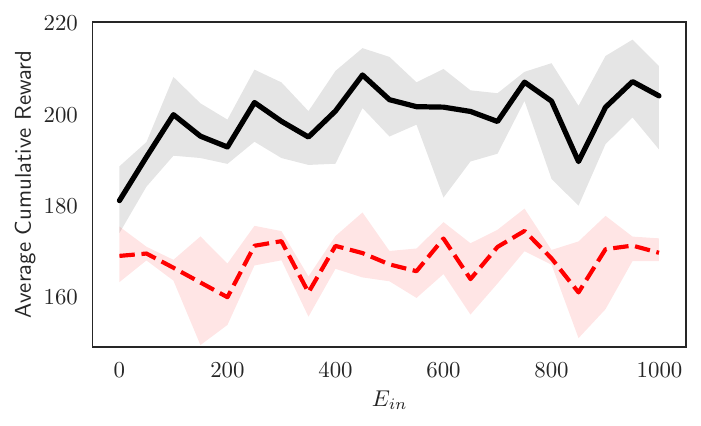}
\end{subfigure}
  \caption{The left panel shows the discrete distributions of the $\eta$ values selected respectively by PPO (in black) and by a Bernoulli distribution (in red) with a parameter equal to 0.5 over $200$ Monte Carlo simulations of the system. The right panel shows the average cumulative fitness of the system as a function of the number of training episodes for the trained PPO instances (in solid black) and the Bernoulli distribution of $\eta$ (in dashed red) with the corresponding confidence intervals.}
\label{Fig:eta_distribution_learning}
\end{figure}

After training the PPO algorithm, the reinforcement learning agents tend to select only the extremes of the set $\mathcal{A}=\{0,0.5,1\}$, which corresponds to an interest rate strategy ($\eta = 0.0$) or a liquidity strategy ($\eta = 1.0$). For this reason, we highlight such a dichotomy in the baseline model since it is the pattern that emerges when all the banks in the system follow the policy recommendation. Considering the emerging dichotomy in the selected action, we compare the PPO performance with respect to a dynamic random baseline that picks the value of $\eta$ according to a Bernoulli distribution with a parameter equal to 0.5. This random policy that chooses between 0 and 1 with equal probability represents a meaningful benchmark, as we observe in the left-hand side of Fig. \ref{Fig:eta_distribution_learning}, where the values of $\eta$ in both scenarios are identically distributed over the $200$ performed Monte Carlo simulations. The Kolmogorov-Smirnoff test statistically confirms up to the $1\%$ confidence level that the distribution of the $\eta$ values generated by the selected\footnote{It is common in reinforcement learning applications to train different instances of the same algorithm and then select the best performing one over some out-of-sample tests \citep{andrychowicz2020matters}} PPO instance is not significantly different from the one of the random baseline. The right-hand side of Fig. \ref{Fig:eta_distribution_learning} summarizes the learning process results where the system's average cumulative fitness in Eq. \ref{Eq:reward} is represented on the $y$-axis. Every PPO instance is tested $E_{out}=5$ times using Monte Carlo simulations of length $T$. We notice that the performance metric is always greater for PPO than the random recommendation, signaling that the banks in the system generated by the PPO signal tend to be more attractive for the borrowers by exhibiting a higher aggregated fitness over time. Moving $\eta$ randomly causes banks to be less attractive to the borrowers in their interbank market. This result implies that the PPO instances learn to choose the value of $\eta$ by leveraging the information available about the system without changing the distribution of the values with respect to the random case. The learning procedure allows us to discover when picking a side in this trade-off is convenient.  A further comparison with a decentralized mechanism for the $\eta$ dynamics is provided in Sec.\ref{Subsec:comparo}.

\FloatBarrier

%https://www.frontiersin.org/articles/10.3389/frai.2021.752558/full SHAP reference
%https://christophm.github.io/interpretable-ml-book/shap.html Molnar book
%https://medium.com/@gabrieltseng/interpreting-complex-models-with-shap-values-1c187db6ec83
In order to shed light on the decisions taken by the best performing trained PPO instance, we use the SHapley Additive exPlanation (SHAP)\footnote{For the implementation, we use the Python package linked to \cite{lundberg2017unified}} framework (see \cite{lundberg2017unified}, \cite{shapley201617}). This approach explains a complex nonlinear model like a neural network by shedding light on the contribution of each input feature to the output formation. For each input vector $x \in \mathbb{R}^K$ and a model $f$, the SHAP value $\phi_i(f,x)$, $i=1,\ldots , K$ quantifies the effect (in a sense, the importance) on the output $f(x)$ of the $i$-th feature. To compute this effect one measures, for any subset  $S\subseteq \{1,\ldots, K\}$, the effect of adding/removing the $i$-th feature to the set, i.e. $f_{S\cup\{i\}}(x)-f_S(x)$. The SHAP value is defined as the weighted average 
\begin{equation}
\phi_{i}(f, x)=\sum_{S\subseteq \{1,\ldots, K\}\setminus \{i\}}\frac{\left|S\right| !\left(K-\left|S\right|-1\right) !}{K !}\left[f_{S\cup\{i\}}(x)-f_S(x)\right],
\end{equation}
where the weights ensure that $\sum_i \phi_i = f(x)$.

\begin{figure}[t]
\centering
\begin{subfigure}{.5\textwidth}
  \centering
  \includegraphics[width=\linewidth]{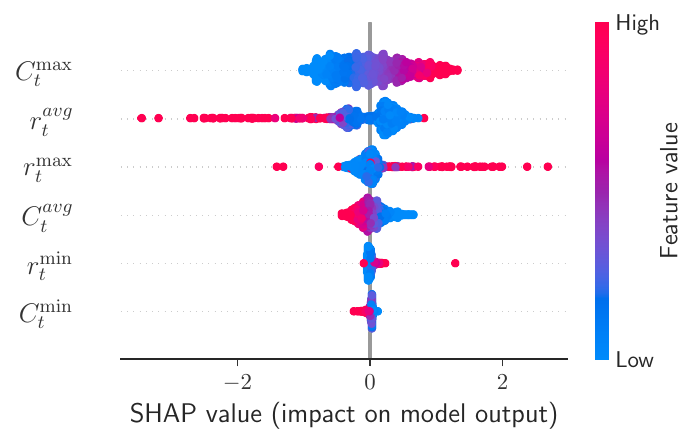}
\end{subfigure}%
\begin{subfigure}{.5\textwidth}
  \centering
  \includegraphics[width=\linewidth]{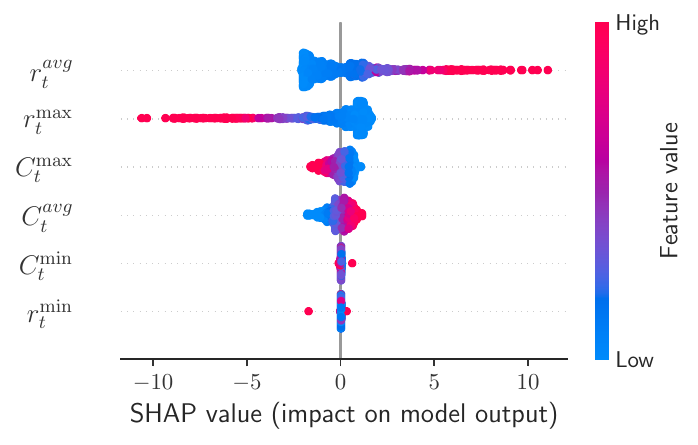}
\end{subfigure}
  \caption{SHAP values relative to the strategy outputs $\eta=0$ (left panel) and $\eta=1$ (right panel). The cloud of colored dots for each input variable expresses the importance and correlation concerning the model output. Features are ordered on the $y$-axis by relevance, so the first on the top influences the model output the most.}
\label{Fig:SHAP}
\end{figure}

Figure \ref{Fig:SHAP} shows the magnitude of the Shapley values for the policy recommendation learned by the best performing PPO instance referred to the two possible outcomes $\eta=0$ and $\eta=1$. The left-hand side shows that high values for the maximum liquidity available in the system tend to favor the choice of an $\eta$ based on the interest rate. Also, a low average interest rate and a high maximum interest rate point to the choice of $\eta=0$. The right-hand side shows an opposite input relevance with a dominant role for high values of the average interest rate and low values of the maximum interest rate. The two figures show that the trained learning algorithm chooses one of the two signals by looking at the main characteristics of the opposite one. When it chooses $\eta=0$, it is more interesting to know if there are participants in the network who are large. In contrast, when it chooses $\eta=1$, it looks for homogeneity of interest rate, a common feature obtained by always playing towards the interest rate. The learning algorithm suggests a switch towards the other competing recommendation to avoid extreme cases in which a disadvantage of one or the other choice exacerbates. For instance, a huge financial institution that gathers all the borrowers' demand when $\eta=1$ could not be sustainable in the long term, so the algorithms suggest switching to the other option. On the other hand, most medium-sized banks offer medium rates when $\eta=0$ cannot gather enough liquidity to deal with deposit shocks, and it would be better to resort to the opposite signal.  

\FloatBarrier

%%%%%%%%%%%%%%%%%%%%%%%%%%%%%%%%%%%%%%%%%%%%%%%%%%%%%%%%%%%%%%%%%%%%%%%%%%%%%%%%%%%%%%%%%%%%%%%%%%
\subsection{Micro and macro consequences of the policy recommendation}\label{Subsec:mmm}
In this subsection, we deal with the implications that the dynamics of the $\eta$ parameter have on the interbank network morphology and the resulting performances of the financial institutions. Finally, we study the effects of the emerging network topology on the market's stability.  All network-related results presented in the following Sessions refer to the active credit network.\\
\subsection*{Topology and evolution of the interbank network}
\begin{figure}
\centering
\begin{subfigure}{.5\textwidth}
  \centering
  \includegraphics[width=\linewidth]{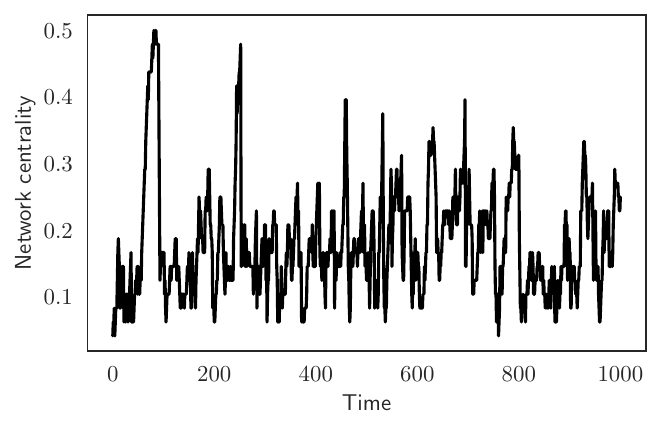}
\end{subfigure}%
\begin{subfigure}{.5\textwidth}
  \centering
  \includegraphics[width=\linewidth]{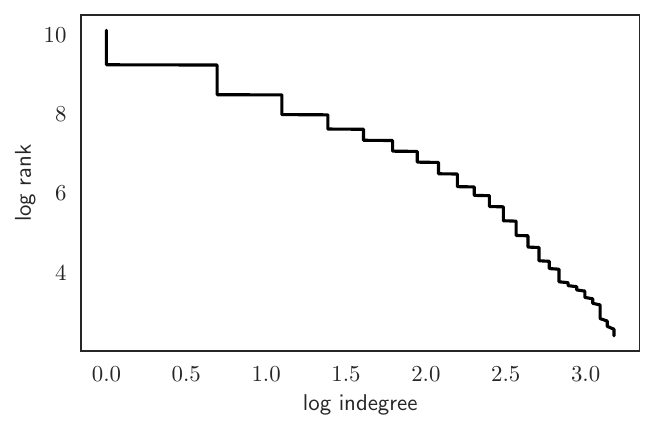}
\end{subfigure}
  \caption{Time series of interbank network centrality (left side). The decumulative distribution function (DDF) of the in-degree (right side).}
\label{Fig:topology}
\end{figure}

Before starting the analysis, it is worth remembering the dynamics of $\eta$, that appear in the banks' fitness (see Eq. \ref{Eq:fitness}), which determines the probability of creating credit links in the system as shown in Eq. \ref{Eq:probattachment}. Therefore, it is appropriate to begin the analysis by describing the topology of the interbank network.  
\begin{figure}
\centering
\begin{subfigure}{.5\textwidth}
  \centering
  \includegraphics[width=\linewidth]{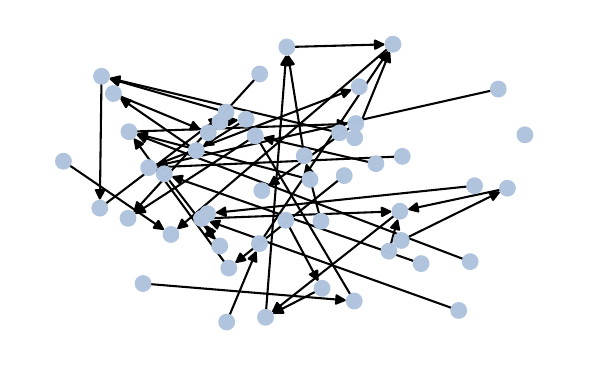}
\end{subfigure}%
\begin{subfigure}{.5\textwidth}
  \centering
  \includegraphics[width=\linewidth]{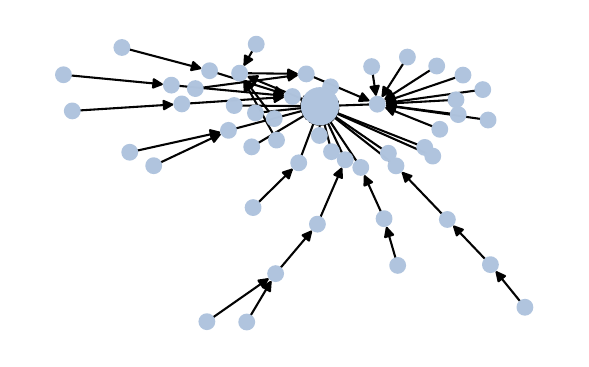}
\end{subfigure}
  \caption{Network configuration at time t=0 (left side), and t=800 (right side).}
\label{Fig:network_topology}
\end{figure}
\FloatBarrier
In Fig. \ref{Fig:network_topology}, we plot the configuration of the endogenous interbank network at two different time steps of a single simulation of the system. As the reader can appreciate, the market configuration goes through different phases ranging from a random topology with isolated agents to a highly centralized architecture where a few hubs compete in credit supply. A more detailed analysis of the evolution of the interbank network architecture over time can be found on the left-hand side of Fig. \ref{Fig:topology}, where we show the time series of network degree centrality 
\begin{equation}
C_{t}^{\text{Net}}=\frac{\sum_{i}\left(k^{\max }_{t}-k^{i}_{t} \right)}{N(N-1)-|V_t|},
\end{equation}
where $N$ is the number of banks, $|V_t|$ is the total number of incoming links in the system, $k^{i}_{t}$ is the number of incoming links for the $i$-th bank, and $k^{\max }_{t}$ is the number of incoming links holds by the hub of the network.

\FloatBarrier
The dynamics of network centrality show how the morphology of the credit market evolves, going from periods in which the network is decentralized and made of many small components to periods in which more than 45\% of banks are connected to a single hub. In addition, the topology of the emerging network is different from that of the random graph, where the in-degree distribution decays exponentially. Similar to real credit networks, in our system, some banks are found to have a disproportionately large number of incoming links. In contrast, others have very few (see \cite{iori2018empirical}, for a survey of the relevant literature). This result is shown in the right-hand side of Fig. \ref{Fig:topology} where we plot the decumulative distribution function of the in-degree. As the reader can observe, this distribution is in keeping with scale-free networks and displays a 'fat tail.'

\begin{figure}[t!]
\begin{minipage}{\textwidth}
  \centering
  \includegraphics[width=0.8\linewidth]{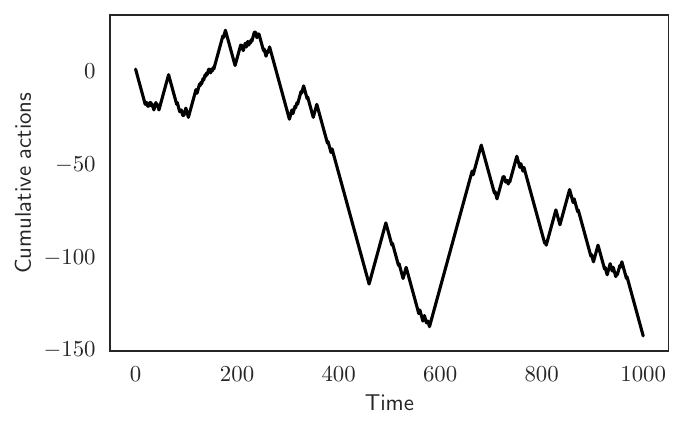}
\end{minipage}%
\vspace{1em}
\begin{minipage}{\textwidth}
    \centering
    \begin{tabular}{l@{\hskip 2em}c@{\hskip 2em}c}
    \toprule
    \textbf{$y_{t}$} & $b_{0}$ & $b_{1}$ \\
    \midrule
    % \multirow{2}{*}{Log diff ID}      &    0.0010 &                                -0.0021 \\[-0.7ex]
    %                           &    {\scriptsize(0.52)}        &                                {\scriptsize(-0.77)} \\[0.8ex]
  %  \multirow{2}{*}{Liquidity}      &  259.49\threeS &                                69.18\threeS \\[-0.7ex]
   %                            &    {\scriptsize(491.70)}        &                                {\scriptsize(51.60)} \\[0.8ex]
%    \multirow{2}{*}{Equity}         &   21.04\threeS &                                -3.19\threeS \\[-0.7ex]
 %                              &    {\scriptsize(667.17)}        &                                {\scriptsize(-11.80)} \\[0.8ex]
  %  \multirow{2}{*}{Incoming links} &    9.78\threeS &                                -0.22\threeS \\[-0.7ex]
   %                            &    {\scriptsize(381.17)}        &                                {\scriptsize(-55.45)} \\[0.8ex]
    \multirow{2}{*}{Centrality} &    0.1830\threeS &                                -0.0047\threeS \\[-0.7ex]
                               &    {\scriptsize(599.06)}        &                                {\scriptsize(-11.8013)} \\[0.8ex]
    \multirow{2}{*}{Density} &    0.1042\threeS &                                -0.0070\threeS \\[-0.7ex]
                               &    {\scriptsize(317.08)}        &                                {\scriptsize(-16.01)} \\[0.8ex]
   % \multirow{2}{*}{Density undir} &    0.2084\threeS &                                -0.0141\threeS \\[-0.7ex]
%                               &    {\scriptsize(317.08)}        &                                {\scriptsize(-16.01)} \\[0.8ex]
    \multirow{2}{*}{Diameter} &    10.14\threeS &                                0.22\threeS \\[-0.7ex]
                               &    {\scriptsize(1104.02)}        &                                {\scriptsize(17.10)} \\[0.8ex]
   % \multirow{2}{*}{Diameter undir} &    11.80\threeS &                                0.24\threeS \\[-0.7ex]
    %                           &    {\scriptsize(1344.06)}        &                                {\scriptsize(19.59)} \\[0.8ex]
    \multirow{2}{*}{Components} &    1.48\threeS &                                0.0095\threeS \\[-0.7ex]
                               &    {\scriptsize(656.45)}        &                                {\scriptsize(3.04)} \\[0.8ex]
    \multirow{2}{*}{Avg nodes per components} &    39.50\threeS &                                -0.21\threeS \\[-0.7ex]
                               &    {\scriptsize(940.20)}        &                                {\scriptsize(-3.57)} \\[0.8ex]                         
    \bottomrule
    \addlinespace[1ex]
    \multicolumn{3}{l}{\textsuperscript{***}$p<0.01$, 
      \textsuperscript{**}$p<0.05$, 
      \textsuperscript{*}$p<0.1$}
    \end{tabular}
\end{minipage}
\vspace*{-1mm}
  \caption{Top Panel: Time series of $\eta$ cumulative values over the simulation. Bottom Panel: Estimated results with the respective T-test in brackets for Eq.\ref{Eq:categorical}. $b_0$ is the estimated mean value of $y$ when $\eta=1$ and $b_1$ the deviation from this mean value when $\eta=0$. Data are obtained through $200$ Monte Carlo simulations of the system.}
\label{fig:cumulo}
\end{figure}
\FloatBarrier
 To conclude the analysis of the interbank market architecture, we deal with the effect of the $\eta$ parameter on the credit network topology. In the top panel of Fig. \ref{fig:cumulo}, we plot a single realization of the cumulative value of $\eta$ over time. The figure shows how the reinforcement learning algorithm generates a time evolution in the choice of policy recommendations. Precisely, increasing (decreasing) values in the curve correspond to a signal that directs the system toward a high liquidity supply (low interest rate), i.e., $\eta=1$ (i.e., $\eta=0$). 
 
The effect of the signal in shaping the topology of the interbank network is, instead, shown in the lower panel of Fig. \ref{fig:cumulo}, where we estimate a categorical regression model
\begin{equation}\label{Eq:categorical}
y_{t} = b_{0} + b_{1}(1-\eta_t),
\end{equation}
where $b_{0}$ is the estimated mean value assumed by the dependent variable $y$ when $\eta=1$ and  $b_{0}+b_{1}$ is the mean when $\eta=0$. As shown in the bottom panel of Fig. \ref{fig:cumulo}, when the system selects low interest rates, the interbank network is less centralized, more sparse, and with a larger diameter. Moreover, the graph is fragmented into many scarcely-populated islands. 

\begin{figure}
\centering
\begin{subfigure}{.5\textwidth}
  \centering
  \includegraphics[width=\linewidth]{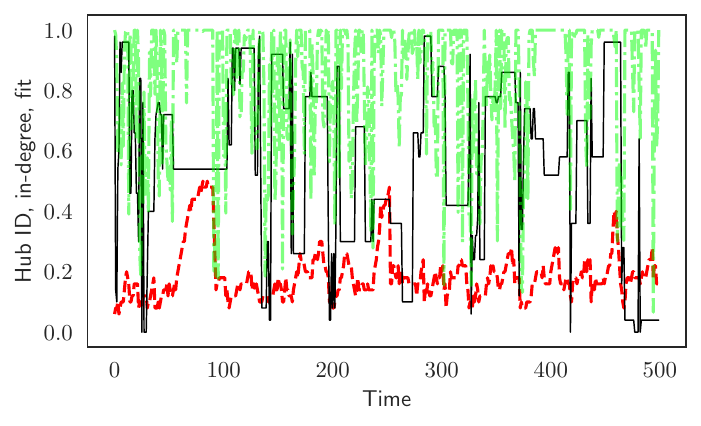}
\end{subfigure}%
\begin{subfigure}{.5\textwidth}
  \centering
  \includegraphics[width=\linewidth]{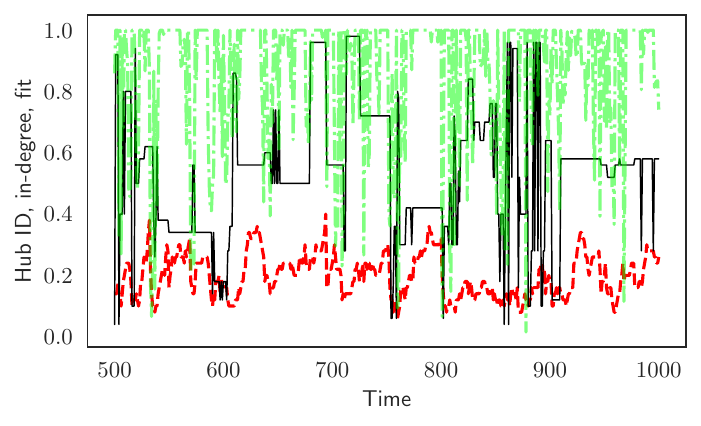}
\end{subfigure}
  \caption{Time series of the most connected lender (hub) evolution along the time T. The solid black line identifies the normalized hub id, the red dashed line her number of clients ( incoming links), and the green dotted line the hub' fitness. Colors are available on the website version.}
\label{Fig:id_link_guru}
\end{figure}

Having described the architecture of the interbank network, let us now examine its evolution over time. It is worth remembering that banks signal in the market their attractiveness $\mu$ according to the recommendation from the regulator, i.e., whether to compete on low interest rates, $\eta=0$, or on high liquidity supply, $\eta=1$. While the regulator's signal is market-specific, liquidity supplies and interest rates (based on  Eq.6 ) are bank-specific variables. This mechanism creates competition among financial institutions for credit allocation. The war in granting credit, modeled through the possibility of redefining lending agreements via Eq. \ref{Eq:probattachment} is shown in Fig. \ref{Fig:id_link_guru}.
\FloatBarrier
The black solid, red dashed, and green dotted lines represent the normalized id of the lender with the highest number of clients (i.e., the hub), her incoming links (i.e., number of clients), and her fitness, respectively. As the reader can appreciate, the simulation presents periods of hub stability and alternation and competition between hubs. When the hub stands out from her competitors and signals a significantly higher fitness (i.e., the green dotted line approaches the unit), she can attract numerous clients, as shown by her high number of incoming links. However, the attractiveness of the hub may work against her. A large portfolio of customers increases the likelihood that some of them may fail. This either decreases the attractiveness of the hub herself or even causes her failure.  The reduction of the hub's fitness due to one of her clients' failure works in the following way. On the one hand, when the fitness uses a strategy based on a low interest rate, the client's approach to the bankruptcy threshold increases the borrower's financial fragility and probability of bankruptcy. Both these effects increase the lending interest rate, making the hub less attractive (see Eq.\ref{Eq:interest}). On the other hand, when $\mu$ moves towards a high liquidity supply, the borrower's bad debt is absorbed by the lender's net worth. The fall in the latter causes a parallel reduction in the hub liquidity, as shown by the balance sheet identity (see Eq. \ref{Eq:balancesheet}). Interestingly, reducing the hub net worth could reduce liquidity higher than proportionally, given the Basel rules on maximum capital and leverage ratio.  In any case, the decrease in the agent's fitness gradually reduces her clients and makes other lenders more attractive. These agents can replace the unsuccessful hub and so become, in turn, the most appealing lenders.\\
\subsection*{Micro consequences of the reinforcement learning policy}
\begin{figure}
\begin{minipage}{\textwidth}
    \centering
    \begin{tabular}{l@{\hskip 2em}c@{\hskip 2em}c}
    \toprule
    \textbf{$y_{t}$} & $b_{0}$ & $b_{1}$ \\
    \midrule
    \multirow{2}{*}{Liquidity} &  2960.34\threeS &                               331.42\threeS \\[-0.7ex]
                               &    {\scriptsize(1062.34)}        &                                {\scriptsize(82.59)} \\[0.8ex]
    \multirow{2}{*}{Equity}    &   888.96\threeS &                              -110.72\threeS \\[-0.7ex]
                               &    {\scriptsize(1171.31)}        &                                {\scriptsize(-135.79)} \\[0.8ex]
    \multirow{2}{*}{Leverage}  &     0.01673\threeS &                                 0.000175\threeS \\[-0.7ex]
                               &    {\scriptsize(435.43)}        &                                {\scriptsize(3.23)} \\[0.8ex]
    \multirow{2}{*}{Rationing} &     0.33\threeS &                                 0.28\threeS \\[-0.7ex]
                               &    {\scriptsize(71.80)}        &                                {\scriptsize(38.31)} \\[0.8ex]
    \multirow{2}{*}{Bad debt}   &    36.05\threeS &                                 2.19\threeS \\[-0.7ex]
                               &    {\scriptsize(446.46)}        &                                {\scriptsize(19.49)} \\[0.8ex]
    \multirow{2}{*}{Failed banks}   &     3.14\threeS &                                 0.35\threeS \\[-0.7ex]
                               &    {\scriptsize(520.17)}        &                                {\scriptsize(41.00)} \\[0.8ex]
    \bottomrule
    \addlinespace[1ex]
    \multicolumn{3}{l}{\textsuperscript{***}$p<0.01$, 
      \textsuperscript{**}$p<0.05$, 
      \textsuperscript{*}$p<0.1$}
    \end{tabular}
\end{minipage}%
\vspace{1em}
\begin{minipage}{\textwidth}
\centering
  \includegraphics[width=0.8\linewidth]{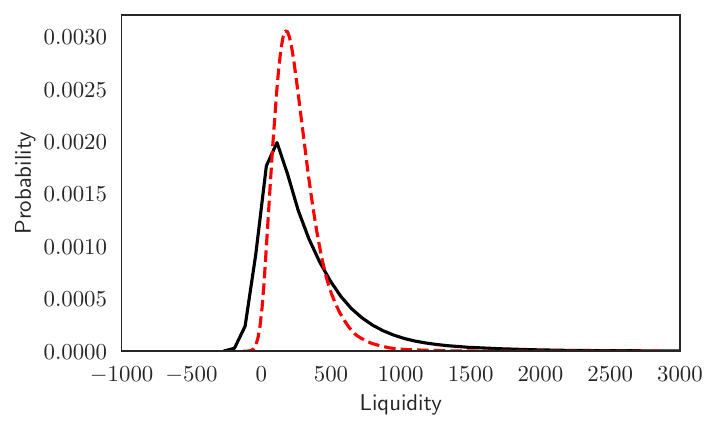}
\end{minipage}
\vspace*{-1mm}
\caption{Top Panel: Estimated results with the respective T-test in brackets for Eq.\ref{Eq:categorical}. $b_0$ is the estimated mean value of $y$ when $\eta=1$ and $b_1$ the deviation from this mean value when $\eta=0$. Data are obtained through $200$ Monte Carlo simulations of the system. Bottom Panel: Density distributions of aggregated liquidity over times over $200$ Monte Carlo simulations. The black solid and red dashed lines show $\eta=0$ and $\eta=1$, respectively.}
\label{Fig:micro_performance}
\end{figure}

In this subsection, we investigate how the dynamics of $\eta$ affect the hub's performance and other financial institutions. 
\begin{figure}
  \centering\includegraphics{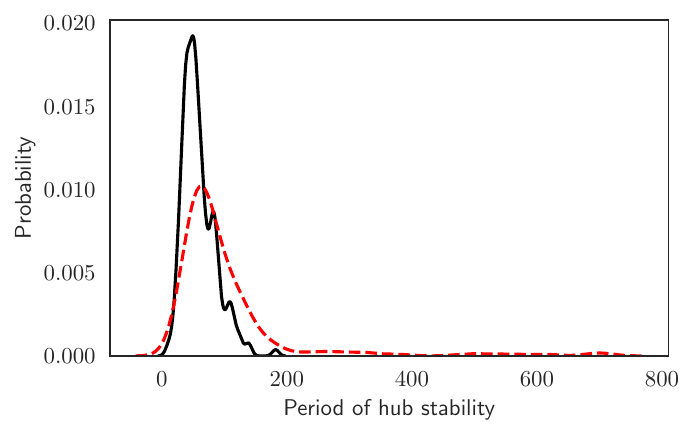}
  \caption{Density distributions over $200$ Monte Carlo simulations of the maximum period of hub stability in which the strategy does not change. The black solid and red dashed lines show $\eta=0$ and $\eta=1$, respectively.}
  \label{Fig:guru_dist}
\end{figure}
\FloatBarrier
In Fig. \ref{Fig:guru_dist}, we show how choosing between a low interest rate and a high liquidity supply strategy affects the hub's longevity. The figure shows the distribution, over $200$ simulations, of the maximum period of hub stability in which the strategy does not change, respectively, for $\eta=0$ (black) or $\eta=1$ (red). The figure shows that the hub is generally more stable if the regulator recommends a high liquidity supply (red dashed line in Fig \ref{Fig:guru_dist}). Moreover, also at a micro level, we show that $\eta=1$ seems to produce better individual performances. This result is shown in the top panel of Fig. \ref{Fig:micro_performance}, where we report the effect of the two possible values of $\eta$ on some key individual variables. 

\FloatBarrier
Specifically, our results, estimated via the categorical regression model in Eq. \ref{Eq:categorical}, show that a signal that directs the system toward an abundant supply of liquidity (i.e., $\eta=1$) produces better results in controlling leverage, rationing, bad debt, and bankruptcies. Moreover, according to the hypothesis that banks fail as net-worth falls below a minimum threshold, the equity is higher in the case of $\eta=1$. \\The result on the liquidity is, however, less intuitive. The system that competes on the interest rate level is significantly more liquid than the one adopting high liquidity, with an average liquidity value of 3291 in the case of $\eta=0$ and 2960 in the opposite case. The reason for the apparent better performance on liquidity in the case of $\eta=0$ lies in the competition among banks using interest rates. As clarified by Eq. \ref{Eq:interest}, the financial institutions applying the lowest interest rates are the smallest ones. This implies that the biggest banks are less attractive to borrowers because they charge higher rates. Therefore, the system excludes these economic agents from trading while encouraging small institutions to provide liquidity. This mechanism of selection has a twofold effect. On the one hand, it generates a substantial unbalance between lenders and borrowers size.
Creditors, much smaller than debtors, are overwhelmed in the event of their clients' bankruptcy. On the other hand, the exclusion from the exchanges of the largest institutions leaves a consistent level of unallocated liquidity in the system. The first effect, i.e., agents' unbalance, determines the worst performances under $\eta=0$, while the second effect, i.e., exclusion, determines the highest level of unallocated liquidity in the system. In contrast, a signal that directs the system towards an abundant liquidity supply produces a more ''homogeneous'' distribution among banks' liquidity, as shown in the bottom panel of Fig. \ref{Fig:micro_performance}. This balance between economic agents generates a uniform risk exposure among counterparties, favoring the system's resiliency in front of shocks.  This result, although not unanimously shared (see \cite{haldane2011systemic}), is in line with other studies showing that the imbalance between lenders and borrowers size is a leading force in generating propagation of systematic failure (see, for instance, \cite{caccioli2012heterogeneity}, \cite{berardi2017banks}, \cite{iori2006systemic}, \cite{lenzu2012systemic} and \cite{tedeschi2012bankruptcy}).
In the language of network theory, the scenario corresponding to $\eta=1$ can be interpreted as an assortative mixing, that is a preference for the network's nodes (banks) to attach to others that are similar in some way (i.e in size in our contest). Symmetrically, $\eta=0$ generates a disassortative mixing, where big-in-size borrowers tend to attach to low-in-size lenders. The presence of a strong disassortative mixing in interbank markets, and specifically of an imbalance between borrowers and lenders is also empirically found in the e-MID interbank market (see, for instance, \cite{de2006fitness} \cite{de2007italian}).
\subsection*{Systemic impact of the network} \label{Subsec:systemicimpact}
\begin{table}[!ht]
\centering
\begin{tabular}{l@{\hskip 2em}c@{\hskip 2em}c@{\hskip 2em}c@{\hskip 2em}}
\toprule
 \textbf{Indep. Variable} & \multicolumn{3}{c}{\textbf{Dep. Variable}} \\
 &   Rationing &  Failed banks & Leverage  \\
\midrule
\multirow{2}{*}{Net centrality} &   -0.25\threeS & -2.09\threeS & -0.016\threeS  \\[-0.7ex]
                           &  {\scriptsize(-6.04)} &  {\scriptsize(-41.82)} &  {\scriptsize(-55.69)} \\[0.8ex]

\multirow{2}{*}{Density} &   -1.32\threeS & -9.14\threeS & -0.051\threeS  \\[-0.7ex]
                           &   {\scriptsize(-52.64)} &  {\scriptsize(-254.08)} &   {\scriptsize(-235.33)} \\[0.8ex]

%\multirow{2}{*}{Density undir} & -0.66\threeS & 	-4.57\threeS & 711.15\threeS & -0.025\threeS  \\[-0.7ex]
 %                          &  {\scriptsize(-52.64)} &  {\scriptsize(-254.08)} &  {\scriptsize(58.03)} &  {\scriptsize(-235.33)} \\[0.8ex]   

\multirow{2}{*}{Diameter} &  0.011\threeS & 0.032\threeS & 0.0002\threeS  \\[-0.7ex]
                           &  {\scriptsize(8.95)} &  {\scriptsize(21.69)} &  {\scriptsize(27.02)} \\[0.8ex]

% \multirow{2}{*}{Diameter undir} & 0.013\threeS & 0.037\threeS & -11.12\threeS & 0.0002\threeS  \\[-0.7ex]
%                           &  {\scriptsize(9.71)} &  {\scriptsize(23.86)} &  {\scriptsize(-13.59)} &  {\scriptsize(27.68)} \\[0.8ex]  

\multirow{2}{*}{Components} &  0.029\threeS & 0.020\threeS & 0.0004\threeS  \\[-0.7ex]
                           &   {\scriptsize(5.28)} &  {\scriptsize(3.09)} &  {\scriptsize(10.57)} \\[0.8ex]

\multirow{2}{*}{Avg nodes per comp} &  -0.0011\threeS & -0.0022\threeS & -0.00002\threeS  \\[-0.7ex]
                           &    {\scriptsize(-4.18)} &  {\scriptsize(-6.97)} &  {\scriptsize(-13.19)} \\[0.8ex]      
                           
\bottomrule
\addlinespace[1ex]
\multicolumn{3}{l}{\textsuperscript{***}$p<0.01$, 
  \textsuperscript{**}$p<0.05$, 
  \textsuperscript{*}$p<0.1$}
\end{tabular}
\vspace{1.5em}
\caption{Regression results between indicators of the interbank stability and network measures. T-stats for each coefficient are provided in parentheses. Data are obtained through $200$ Monte Carlo simulations of the system.}
\label{reg}
\end{table}
To conclude the section, we combine the results on network topology and individual performance as a function of $\eta$ to capture the interbank architecture's overall effect on systemic stability. To this end, in Tab. \ref{reg}, we report the results of a linear regression estimated through ordinary least squares where the independent variables are some measures of the interbank network topology and dependent variables are some indicators of systemic market stability. In line with what has been observed so far, when the network tends to be centralized, i.e., denser towards the hub and with a smaller diameter, the risk of contagion decreases, i.e., bankruptcies, rationing, and leverage are reduced. This architecture corresponds to a graph composed of a few highly populated components. It is worth noting that this topology emerges when the interbank system is oriented towards an abundant supply of liquidity, which generates a certain homogeneity among agents able to compensate for the imbalance between lenders and borrowers present in the case of $\eta=0$. In this respect, clarification is essential: $\eta=1$ is not the absolute best signal. This is the best strategy given the individual and aggregate conditions of the system at the time of the choice. The algorithm is designed to identify one recommendation as optimal based on the underlying environmental conditions. The robustness of this observation is shown in Sec.\ref{Subsec:macro} and Sec. \ref{Subsec:comparo}. In the former, we show that the system governed by a regulator that directs the choice via the implemented reinforcement learning algorithm outperforms a system based on a random selection between the two signals. In the latter, we demonstrate the better performances of the reinforcement learning rather than modeling an $\eta$ evolving with decentralized dynamics.
\subsection{The reinforcement learning based recommendation for taming systemic risk}\label{Subsec:macro}
In this subsection, we study the effect on the interbank systemic stability of the policy recommendation obtained through the reinforcement learning mechanism solved by the PPO algorithm. 

%Inspired by the former governor's words, in this subsection, we use the $\eta$ variable as a policy recommendation, obtained by a reinforcement learning mechanism solved with the PPO algorithm, to analyze its impact on the interbank market (in)stability.  
%investigate the effect of a policy strategy, obtained with a reinforcement learning mechanism solved via the PPO algorithm, on the interbank market (in)stability.
Specifically, we answer the following question: how would the interbank system perform in terms of aggregate resiliency when the regulator directs financial institutions to choose the optimal strategy between competing on the low interest rate, $\eta=0$, or on high liquidity, $\eta=1$? Again we compare the effects of the learned strategy on the market stability with those of a random strategy. Finally, the last part of this Section is devoted to understanding the market performance as the percentage of banks that follow the policy recommendation changes.

Before delving into the analysis of interbank market stability as a result of the policy recommendation, it is appropriate to clarify how contagion develops and propagates in the model. 

\begin{figure}[h]
\centering
\begin{subfigure}{0.5\textwidth}
\centering
  \includegraphics[width=0.75\linewidth]{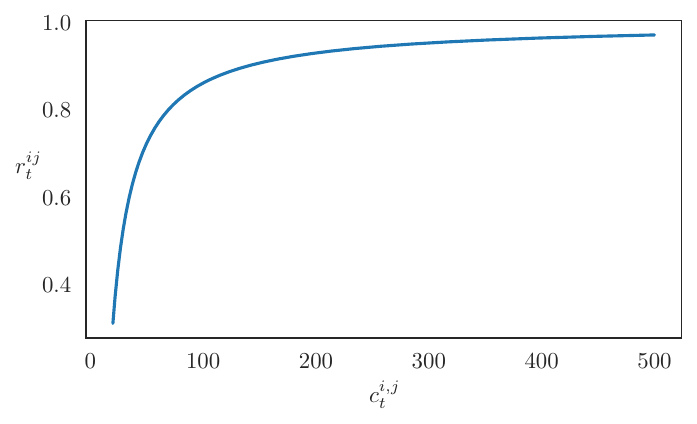}
  \end{subfigure}%
  \begin{subfigure}{0.5\textwidth}
  \centering
  \includegraphics[width=0.75\linewidth]{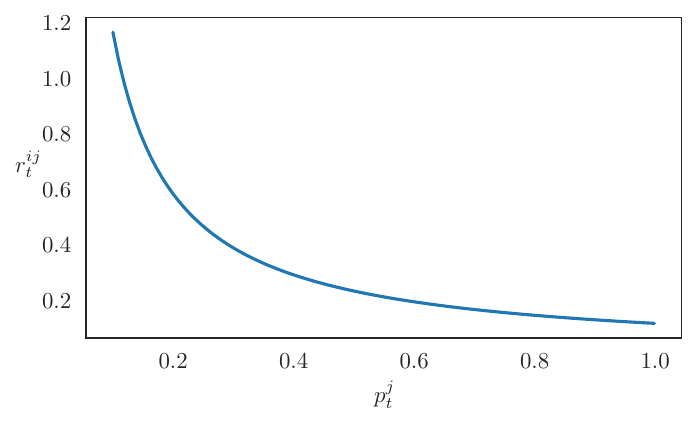}
  \end{subfigure}%
\vspace*{-5mm}
  \caption{Interest rate as a function of the lending capacity $c_t^{i,j}$ and the probability of surviving $p_t^{j}$.}
  \label{Fig:decum_liquidity_ir}
\end{figure}
\FloatBarrier
When a bank is hit by a negative shock and is unable to cope with it through her own resources or those obtained in the interbank market, she fails and leaves the economic system. However, her death has important systemic repercussions. In fact, she may generate a bad debt to her lender in the financial market. As can be seen from Eq.\ref{Eq:equity_motion}, the bad debt of the borrower propagates to her lender through a lowering of the creditor's equity. On the one hand, a sufficiently large bad debt could directly bankrupt the lender and thus generate a new failure. On the other hand, our lender would be, in any case, weakened, with some important financial consequences, namely an increase in her probability of bankruptcy and in her leverage and a decrease in her capacity, i.e. the maximum amount of credit she would obtain in case of need in the interbank market. These three ingredients worsen our creditor’s credit conditions. Therefore, in the unfortunate event of a negative shock hitting this agent with her concomitant need for liquidity in the interbank market, her current credit conditions would provide the bank with a considerably higher interest rate and a lower granted loan. In fact, as it can be seen from the numerical study of the interest rate equation (Eq.\ref{Eq:interest}), this function is positive in capacity and negative in the bank surviving probability (see Fig.\ref{Fig:decum_liquidity_ir}). Of course, this could lead to the failure of our agent and the concomitant infection of her lender.

In order to prove the real ability of the model to generate contagions, we study if there are bankruptcy cascades in the artificial system. Since the purpose of this exercise is to study the evolution of a self-contained system with a given initial number of banks, we exclude the possibility that failing banks would be replaced by new entrants.  
\begin{figure}
 \centering\includegraphics{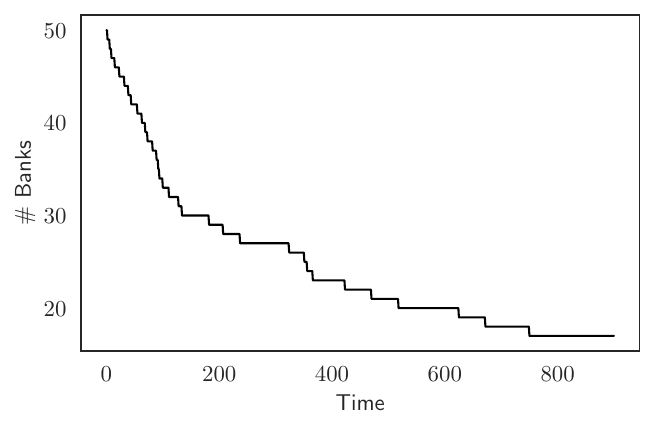}
 \vspace*{-5mm}
 \caption{Average number of surviving banks as a function of time. Results are obtained  without reintroducing the failed agents and running 10 Monte-Carlo simulations.}
 \label{Fig:cascades}
\end{figure}
 \FloatBarrier
Fig. \ref{Fig:cascades} displays the average number, over 10 Monte-Carlo simulations, of surviving banks as a function of time. The slope of the number of surviving banks curve provides evidence of contagious failures, that is periods in which many banks collapse together (see \cite{iori2006systemic} and \cite{tedeschi2012bankruptcy}, for similar results).
We can conclude that the default of an agent may increase systemic risk in our framework. In fact, our dynamics not only generate bankruptcies but also a rapid decline in the time path of surviving banks over time.

Let us now return to the main assumption of this Section, namely the effect on the system resiliency of the policy recommendation obtained via the reinforcement-learning algorithm and its comparison with the random strategy.
 
A common finding in several theoretical and empirical works is that the interbank market works better when credit flows efficiently through the system, thus ensuring it against liquidity shocks (see, for instance, \cite{allen2000financial,freixas2000systemic,carlin2007episodic}).

\begin{figure}
\centering
\begin{subfigure}{.5\textwidth}
  \centering
  \includegraphics[width=\linewidth]{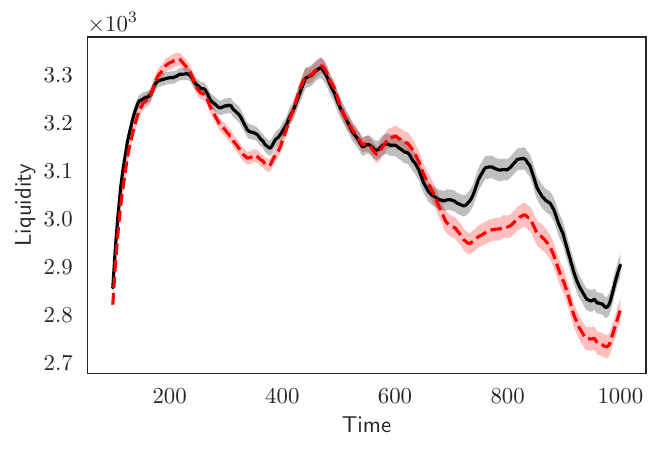}
\end{subfigure}%
\begin{subfigure}{.5\textwidth}
  \centering
  \includegraphics[width=\linewidth]{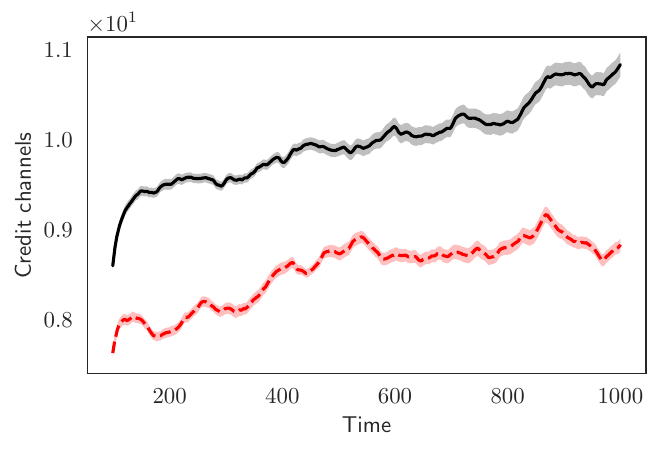}
\end{subfigure}
 \caption{Liquidity of the system (left panel) and the number of credit channels (right panel). Black solid and red dashed lines refer to the best-performing reinforcement learning optimal and random strategies, respectively. The curves reproduce the mean and the standard deviation over 200 simulations of the system and a rolling window of 100 timesteps.}
  \label{Fig:macro_liquidity}
\end{figure}%
\FloatBarrier

Starting from this consideration and recalling the severity of liquidity crises, we show in Fig. \ref{Fig:macro_liquidity} (left side) the effectiveness of the implemented reinforcement learning strategy in spreading liquidity through the system. In the figure, once the best performing learned strategy is selected, as shown on the right-hand side of Fig. \ref{Fig:eta_distribution_learning}, the aggregated average liquidity of 200 simulations over a rolling window of 100 timesteps is shown through time. Although the learned strategy strongly competes with the random one in some periods, its supremacy becomes evident from step 700 onwards. In addition, the average liquidity, over all periods and simulations, of the learned strategy is statistically higher than the one obtained with the random strategy (i.e., 3129.98 (std. 1.5128) vs. 3091.51 (std. 4.4258), respectively).

\begin{figure}
\centering
\begin{subfigure}{.5\textwidth}
  \centering
  \includegraphics[width=\linewidth]{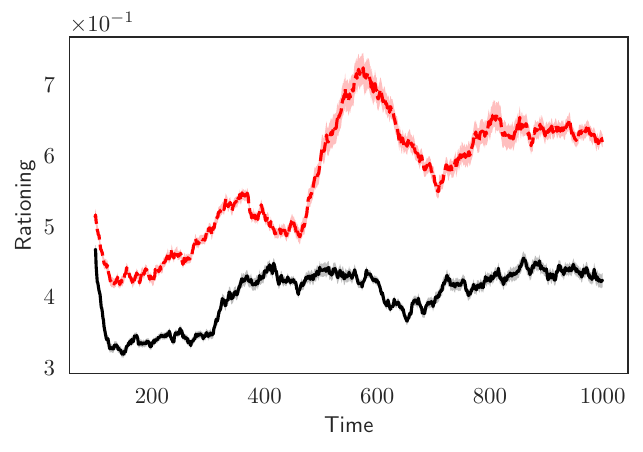}
\end{subfigure}%
\begin{subfigure}{.5\textwidth}
  \centering
  \includegraphics[width=\linewidth]{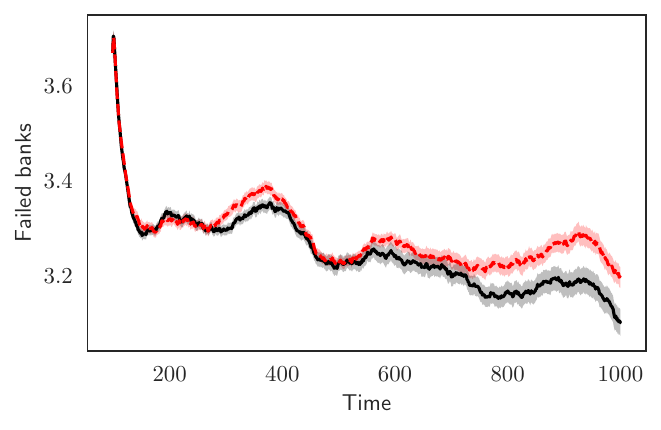}
\end{subfigure}
\caption{Rationing of the system (left panel) and the number of failed banks (right panel). Black solid and red dashed lines refer to the best-performing reinforcement learning optimal and random strategies, respectively. The curves reproduce the mean and the standard deviation over 200 simulations of the system and a rolling window of 100 timesteps.}
\label{Fig:macro_rationing}
\end{figure}

A possible explanation for this phenomenon can be seen in the right-hand side of Fig. \ref{Fig:macro_liquidity}, where we plot the active credit links in the two frameworks\footnote{ By the terms credit channels and credit links we refer to the linkages through which the liquidity needed by borrowers due to the deposit shock flows. These are, therefore, the credit lines used in the active credit network.}. As the reader can appreciate, the number of activated credit channels is higher when the system follows the learned strategy with respect to the case of random strategy, and this guarantees a higher circulation of liquidity in the system. In detail, the average number of credit channels in the first scenario, over time and simulations, is 9.9823 (std. 0.4321), while in the second case is 8.5464 (std. 0.3596). On the whole, this result reveals the ability of the reinforcement learning optimal policy to design an interbank network architecture promoting an efficient credit allocation and, therefore, reducing liquidity shortage phenomena. Consequently, the emerging topology of the credit network effectively controls rationing and avoids failures due to credit crunch phenomena, as shown in Fig. \ref{Fig:macro_rationing}, left and right panels, respectively.
\FloatBarrier
The average values of these variables over all timesteps and simulations confirm the robustness of the two latter results. Specifically, the mean and standard deviation of the rationing in the case of the learned strategy (resp. random strategy) are 0.4024 and 0.0375 (resp. 0.5671 and 0.08465), while the mean and standard deviation of the number of failed banks in the case of the learned policy (resp. random policy) are 3.2101 and 0.0410 (resp. 3.2931 and 0.0423).

It is essential to note the ability of the reinforcement learning mechanism to generate an interbank network whose architecture is resilient in the face of financial attacks. This characteristic provides, on the one hand, an additional monetary policy tool that can be implemented in times of economic adversity and, on the other hand, enriches the vast literature that emphasizes the importance of credit network architecture in dealing with systemic shocks (see \cite{grilli2017networked}, for a survey of the relevant literature).

\begin{figure}
\centering
\begin{subfigure}{.5\textwidth}
  \centering
  \includegraphics[width=\linewidth]{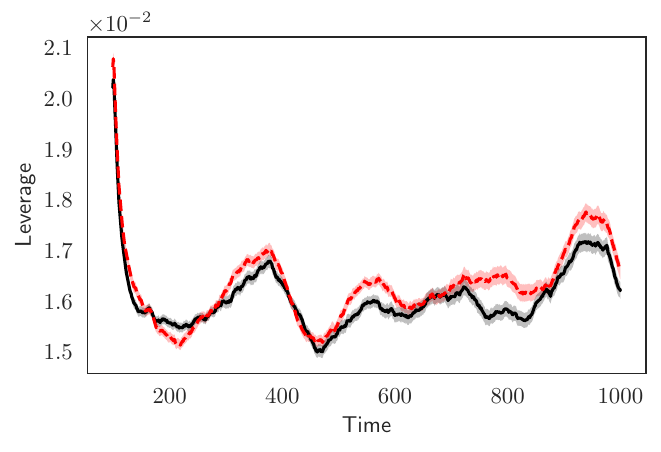}
\end{subfigure}%
\begin{subfigure}{.5\textwidth}
  \centering
  \includegraphics[width=\linewidth]{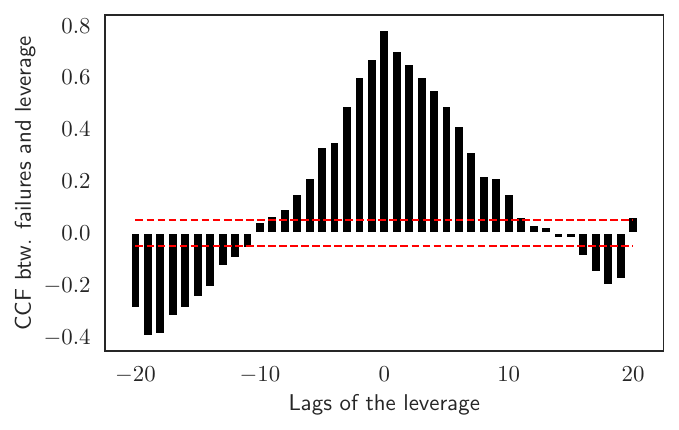}
\end{subfigure}
\caption{Left side: Leverage of the system. Black solid and red dashed lines refer to the best-performing reinforcement learning optimal strategy and to the random strategy, respectively. The curves reproduce the mean and the standard deviation over 200 simulations of the system and a rolling window of 100 time steps. Right side: Average correlation between the number of bankruptcies and lagged leverage, at a 1\% confidence level.}
\label{Fig:macro_leverage}
\end{figure}

We conclude this section by analyzing the effect of the reinforcement learning optimal policy on the market's financial (in)stability. The approach followed here in explaining the materialization of financial frictions is very close in spirit to the Minskyan financial instability hypothesis and therefore uses banks' leverage as the leading indicator (see \cite{minsky1964longer}). In our stylized market, leverage and systemic instability are connected through a specific structure. Given our naive banks' balance sheet (see Eq.\ref{Eq:balancesheet}), leverage is defined as assets on equity. Moreover, credit costs (i.e., interest rates) are strongly positively affected by the leverage (see Eq.\ref{Eq:interest}). When a lender grants a loan to a bank with a low probability of surviving (i.e., an over-leveraged borrower), she charges a higher interest rate via the financial accelerator. This, in turn, exacerbates the borrower's financial condition, pushing her toward a bankruptcy state. If one or more borrowers cannot pay back their loans, even the lenders' equity is affected by bad debts. Therefore, lenders decrease their credit supply and increase the borrowers' rationing. In this way, the profit margin of borrowers decreases, and a new round of failures may occur.
The leverage dynamics when the system follows the reinforcement learning recommended policy and in the random case are shown on the left-hand side of Fig. \ref{Fig:macro_leverage}. The figure highlights two important features. First, the recommended learned policy keeps the leverage below the values obtained with the random policy. Specifically, the average leverage in the first scenario, over time and simulations, is 1.59 (std. 0.042), while in the second case is 1.69 (std. 0.031). Second, the leverage fluctuates over time, thus recalling the different phases of lending suggested by Minsky. There are periods when financial institutions grant more loans without considering the overall financial fragility. However, banks can underestimate their credit risk, making the system more vulnerable when default materializes. This ambiguous effect of the leverage, first positive and then negative, on interbank stability is clearly shown in the right-hand side of Fig. \ref{Fig:macro_leverage}, where the correlation wave between bankruptcies and agents' leverage first decreases from lag $\tau= -21$ up to $\tau= -11$ , then increases from $\tau=-8$ up to $\tau= 9$ , and finally, returns to decrease from $\tau=15$. 
\FloatBarrier

\subsection*{Should they follow or should they not? An exercise on the signal diffusion}\label{herd} 
\begin{figure}
 \centering\includegraphics{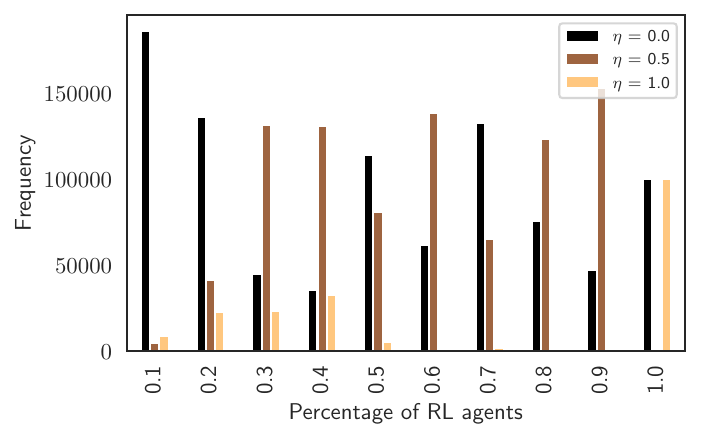}
 \vspace*{-5mm}
 \caption{Discrete distributions of the $\eta$ values selected by PPO over 200 Monte Carlo simulations of the system.}
 \label{Fig:action_bar_mixed}
\end{figure}

 Let us introduce an additional element of heterogeneity concerning the signal itself. Whereas in the previous experiment, all the banks followed the signal on the optimal strategy, here we modify this assumption. We simulate a system where different percentages of banks follow the signal while the others randomly go to different possible strategies. This experiment allows us, on the one hand, to introduce an additional element of differentiation and, on the other, to understand the minimum threshold of followers required by the system with the RL-generated signal to be more resilient than the one with a random strategy. To this end, we fix a percentage $\kappa$ of banks that follow the reinforcement learning strategy, while the remaining $N(1-\kappa)$ banks randomly sample the strategy in the set $\{0,0.5,1\}$. We train the reinforcement learning algorithm following the same procedure as the previous subsections, letting the mixture parameter vary on a discrete range of values. Every time we change the value of $\kappa$, a new algorithm is trained. Several system simulations are performed to evaluate the effect of such heterogeneity in the strategy followed by the banks.\\ Before studying the impact on the interbank systemic stability of the different percentages of financial institutions applying the policy recommendation obtained through the reinforcement learning mechanism and comparing it with the random strategy, an important consideration is necessary. Fig. \ref{Fig:action_bar_mixed} shows that as the rate of followers varies, the PPO algorithm selects different categories of strategies. For example, when only 10\% of the banks follow the optimal signal, the most common strategy steers the banks towards a low interest rate (see solid black line).
However, in this scenario, even if with low probability, a mixed strategy (i.e., $\eta=0.5$) or a high liquidity supply strategy (i.e., $\eta=1.0$) can emerge (see the brown and yellow lines, respectively). This competition among different optimally selected strategies varies as the percentage of followers varies. However, in the case of total synchronization, i.e., when all banks follow the policy recommendation, the system stabilizes, with equal probability, on the two extremes, i.e., $\eta=0$ and $\eta=1$. When $\kappa=0.1$ (i.e., followers are 10\%), the probability of a low interest rate signal is 93\%, while the probability of a mixed strategy is 2.51\%. Instead, the probability of a signal pointing to a high supply of liquidity is 4.38\%. The selected strategies vary when moving towards a percentage of followers of 50\% . Specifically with $\kappa=0.5$ the probability of $\eta=0.0$ is 57\%, that of $\eta=0.5$ is 40\% and finally $\eta=1$ is 3\%. Tab.\ref{tab_prob} shows the portion of the chosen optimal strategy for each percentage of followers.

\begin{figure}[t]
\centering
\begin{subfigure}{.33\textwidth}
  \centering
  \includegraphics[width=\linewidth]{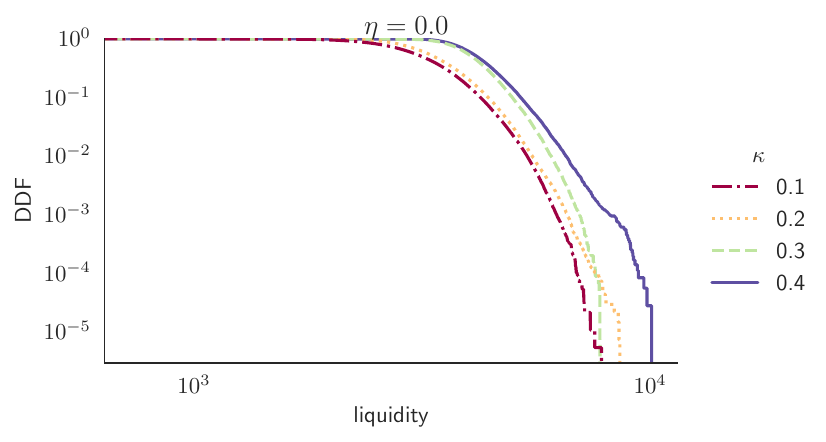}
\end{subfigure}%
\begin{subfigure}{.33\textwidth}
  \centering
  \includegraphics[width=\linewidth]{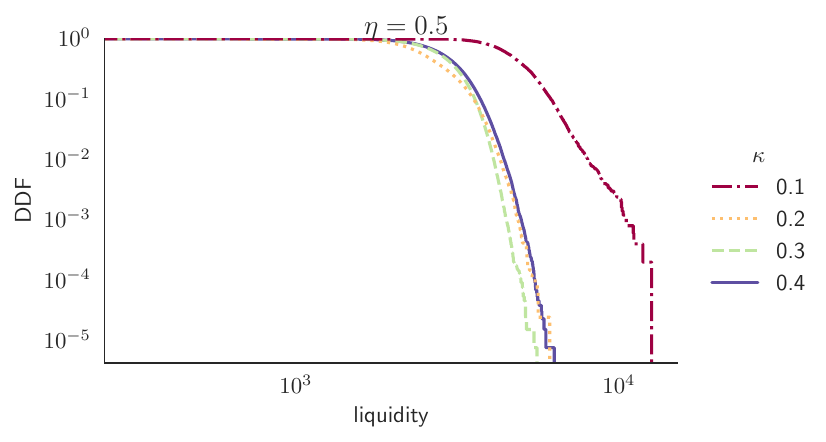}
\end{subfigure}
\centering
\begin{subfigure}{.33\textwidth}
  \centering
  \includegraphics[width=\linewidth]{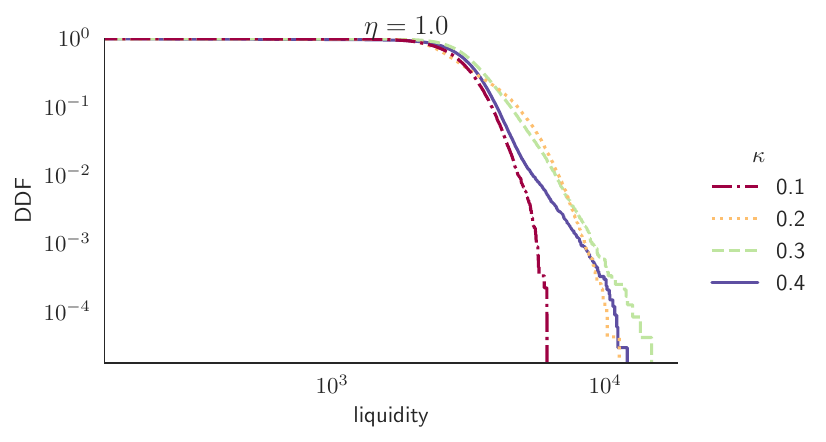}
\end{subfigure}%

\begin{subfigure}{.33\textwidth}
  \centering
  \includegraphics[width=\linewidth]{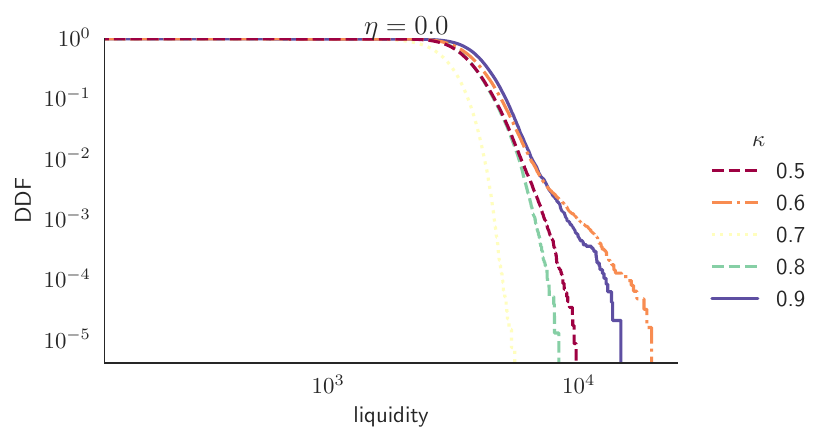}
\end{subfigure}%
\begin{subfigure}{.33\textwidth}
  \centering
  \includegraphics[width=\linewidth]{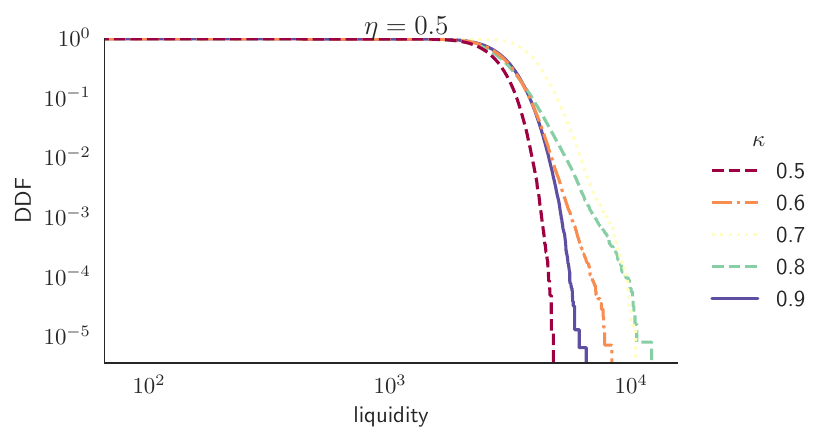}
\end{subfigure}
\centering
\begin{subfigure}{.33\textwidth}
  \centering
  \includegraphics[width=\linewidth]{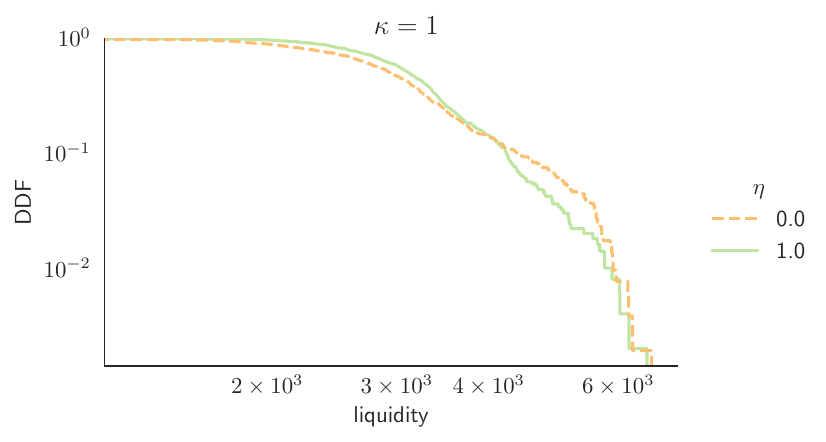}
\end{subfigure}%
  \caption{Decumulative distribution functions for the aggregated liquidity over 200 simulations of the system. The first line displays the distribution corresponding to the emerging strategies (i.e., $\eta=0.0$, $\eta=0.5$ and $\eta=1.0$, first, second and third columns, respectively) with respect to a different percentage of followers ($\kappa$ from 0.1 to 0.4). The second line displays the same distribution for emerging strategy, i.e., $\eta=0.0$ and $\eta=0.5$ first and second column, respectively, with $\kappa$ from 0.1 to 0.9. The third column of the second line reproduces the same distribution for emerging strategy, i.e., $\eta=0.0$ and $\eta=1.0$ when $\kappa=1.0$.}
  \label{Fig:decum_liquidity}
\end{figure}

\begin{table}
    \centering

\begin{tabular}{ll|rrrrrrrrrr}
\toprule
& & \multicolumn{10}{c}{\footnotesize \textbf{Followers percentage}}\\
{} & &   0.1 &   0.2 &   0.3 &   0.4 &   0.5 &   0.6 &   0.7 &   0.8 &   0.9 &  1.0 \\
\midrule
\multirow{3}{*}{\footnotesize \rotatebox[origin=c]{90}{\textbf{Strategy}}}& $\eta$ = 0.0 &  0.93 &  0.68 &  0.22 &  0.18 &  0.57 &  0.31 &  0.66 &  0.38 &  0.24 &  0.5 \\
& $\eta$ = 0.5 &  0.03 &  0.21 &  0.66 &  0.66 &  0.40 &  0.69 &  0.32 &  0.62 &  0.76 &  0.0 \\
& $\eta$ = 1.0 &  0.04 &  0.11 &  0.12 &  0.16 &  0.03 &  0.00 &  0.01 &  0.00 &  0.00 &  0.5 \\
\bottomrule
\end{tabular}
\caption{Average percentage of the chosen optimal strategy ($\eta=0$; $\eta=0.5$ and $\eta=1$) by varying the followers percentage $\kappa$ from 1\% to 100\%. Results are obtained over 200 Monte Carlo simulations of the system}
\label{tab_prob}
\end{table}
 \FloatBarrier
Interestingly, as it can be seen from Tab. \ref{tab_prob}, as $\kappa$ varies, three scenarios emerge. Looking at Tab. \ref{tab_prob}, three scenarios emerge as $\kappa$ varies. For $\kappa$ between 10\% and 40\%, the dynamics of $\eta$ are linear, with $\eta=0.0$ gradually losing the predominance in favor of $\eta=0.5$ and $\eta=1.0$. For $\kappa$ between 50\% and 90\%, an oscillatory dynamic emerges, with $\eta=0.0$ and $\eta=0.5$ alternating continuously. Finally, when the maximum number of followers is reached, we observe an equal distribution between the two pure strategies.  Let us try to understand why these three scenarios emerge.  To this end, in Fig.\ref{Fig:decum_liquidity}, we show the decumulative distribution function of the liquidity in each of the three scenarios, i.e., $\kappa \leq 40\%$, $50\% \leq\kappa \leq 90\%$ and $\kappa=100\%$ for each emerging strategy.  Specifically, the first line of the Fig. \ref{Fig:decum_liquidity} shows the distributions for each value of $\kappa \leq 40\%$ of the three emerging strategies, $\eta=0.0$, $\eta=0.5$ and $\eta=1.0$ first, second and third column, respectively.  The second line displays the same distributions for each value of $50\% \leq\kappa \leq 90\%$ of the two emerging strategies, $\eta=0.0$, $\eta=0.5$, first and second column, respectively.  Finally, the third column of the second line shows the same distribution for $\kappa=100\%$ of $\eta=0.0$ and $\eta=1.0$.

In the first situation, when $10\% \leq\kappa \leq 40\%$, as the percentage of followers increases, what emerges within the three $\eta$ can be summarised as follows.  The decrease in the low-interest-rate strategy depends on the increasing average and heterogeneity of liquidity as $\kappa$ increases.  This is evident in the top left panel of Fig.\ref{Fig:decum_liquidity}, whereas $\kappa$ increases, there is a leftward shift in the liquidity distribution.  The increase in the mixed strategy depends on a more homogeneous distribution of liquidity (and interest rates) and low average values of these two variables (see the top center panel of Fig.\ref{Fig:decum_liquidity}).  Symmetrically with respect to the case of $\eta=0$, the increase in the strategy based on high liquidity is motivated by the increasing heterogeneity in the liquidity distribution and a higher average value of the liquidity as $\kappa$ increases (see top right panel of Fig.\ref{Fig:decum_liquidity}).  In the second scenario, the strong competition and alternation between $\eta=0$ and $\eta=0.5$ depend on an alternation between higher or lower liquidity depending on the prevalence of the mixed or the low interest rate strategy (see the first and second panels at the bottom of Fig.\ref{Fig:decum_liquidity} ). Finally, in the last scenario, where the percentage of followers reaches its maximum (i.e., $\kappa=1$), the system stabilizes, and an equal distribution between the two pure strategies emerges.  In this case, the liquidity distributions of both $\eta$ are heterogeneous, as shown in the last panel to the right of Fig.\ref{Fig:decum_liquidity}.

\FloatBarrier

% \begin{figure}[t]
% \centering
% \begin{subfigure}{0.5\textwidth}
% \centering
%   \includegraphics[width=0.75\linewidth]{liquidity_kappa1_log.pdf}
%   \end{subfigure}%
%   \begin{subfigure}{0.5\textwidth}
%   \centering
%   \includegraphics[width=0.75\linewidth]{interest_eta_kappa1_log.pdf}
%   \end{subfigure}%
% \vspace*{-5mm}
%   \caption{Decumulative distribution functions for the aggregated liquidity and the average interest rate over 200 simulations of the system when all market participants follow the reinforcement learning strategy.}
%   \label{Fig:decum_liquidity_ir}
% \end{figure}
% \FloatBarrier
In summary, the emergence of a strategy or the switching among can be explained as follows. For one of the pure strategies to dominate, the distribution corresponding to the variable representing it must be heterogeneous. Conversely, the emergence of a mixed strategy corresponds to homogeneity in the distribution of both variables, i.e., interest rates and liquidity.

Let us now analyze how the interbank system performs in terms of aggregate resiliency when the regulator convinces different percentages of banks to follow the optimal signal. As in the first part of this subsection, the results obtained with the optimized strategy are compared with those obtained from a random choice of strategy. As in the baseline case, the dynamic random scenario picks the value of $\eta$ according to the probabilities shown in Tab \ref{tab_prob}.

\begin{figure}
\centering
\begin{subfigure}{0.5\textwidth}
  \centering
  \includegraphics[width=\linewidth]{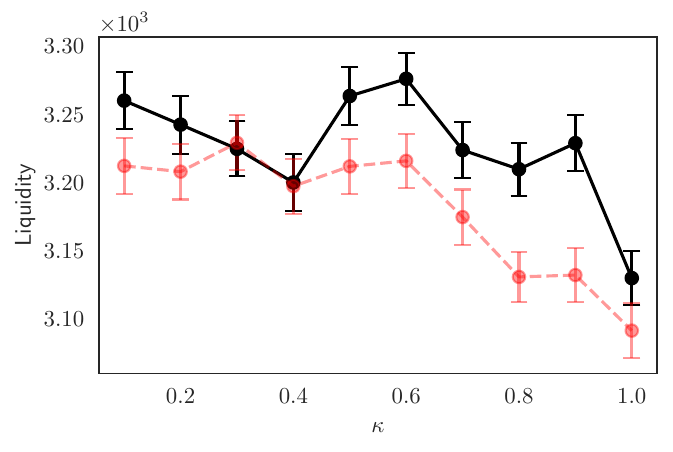}
\end{subfigure}%
\begin{subfigure}{0.5\textwidth}
  \centering
  \includegraphics[width=\linewidth]{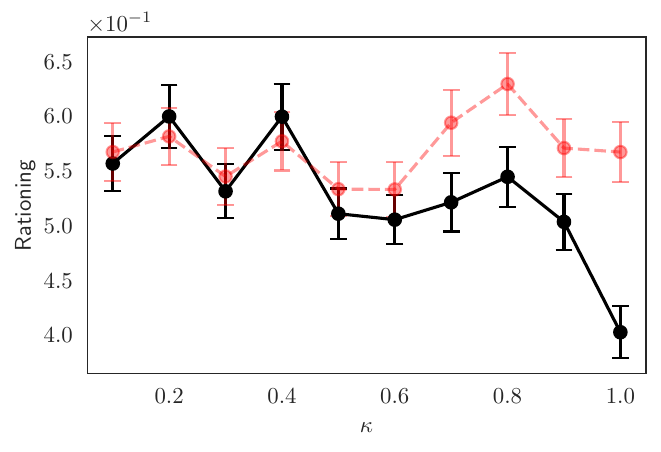}
\end{subfigure}
 \caption{Average liquidity (left panel) and rationing of the system (right panel) as a function of the followers' percentage, $\kappa$. Black solid and red dashed lines refer to the best-performing reinforcement learning optimal and random strategies, respectively. The curves reproduce the mean and the standard deviation over 200 simulations of the system.}
  \label{Fig:mixed_liquidity_leverage}
\end{figure}

\begin{figure}
\centering
\begin{subfigure}{0.5\textwidth}
  \centering
  \includegraphics[width=\linewidth]{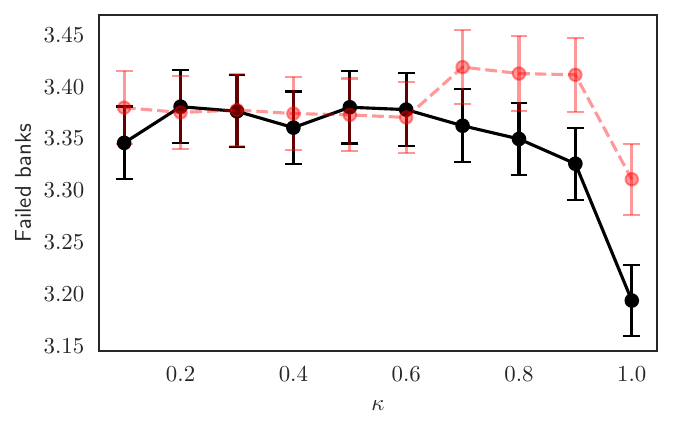}
\end{subfigure}%
\begin{subfigure}{0.5\textwidth}
  \centering
  \includegraphics[width=\linewidth]{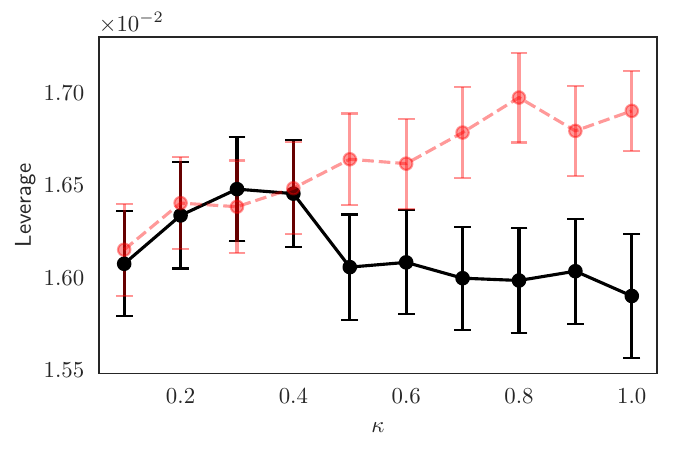}
\end{subfigure}
 \caption{Average number of failed banks (left panel) and their leverage (right panel) as a function of the followers' percentage, $\kappa$. Black solid and red dashed lines refer to the best-performing reinforcement learning optimal and random strategies, respectively. The curves reproduce the mean and the standard deviation over 200 simulations.}
  \label{Fig:mixed_rationing_failed}
\end{figure}

 We report the aggregated results at the macroeconomic level for some of the critical systemic variables. In each panel of Figure \ref{Fig:mixed_liquidity_leverage} and \ref{Fig:mixed_rationing_failed}, we show the variation of the aggregated measure obtained averaging through 200 simulations and through the timesteps of the simulations (1000). The aggregated measure is displayed on the $y$-axis, while the followers' percentage $\kappa$ varies on the $x$-axis.
\FloatBarrier

\FloatBarrier
When the regulator cannot convince a sufficient percentage of banks to follow the policy recommendation, the system generated with the optimal signal obtained via the reinforcement learning algorithm (black solid) does not significantly differ from that generated with the random signal (red dashed lines). This holds for all the considered variables, such as the liquidity and rationing of the system (see Fig. \ref{Fig:mixed_liquidity_leverage}) and the number of failures and leverage of financial institutions (see Fig. \ref{Fig:mixed_rationing_failed}). Instead, when the regulator can convince a share of banks equal to/greater than 60\%, higher systemic stability is observed in the model using the optimized signal than in the random one. In fact, above this percentage, the optimized system, on the one hand, generate higher liquidity and lower rationing, on the other hand, fewer bankruptcies and less leverage for financial institutions.

\subsection{A competition with a decentralized strategy}\label{Subsec:comparo}
In the previous Sections, we compared the aggregate performances of the reinforcement learning strategy with a random strategy that picks the value of $\eta$ according to a Bernoulli distribution with a parameter equal to 0.5. In this Section, we make a comparison with a strategy that selects the $\eta$ parameter in a dynamic and decentralized way so that each bank has her individual plan of action.
Denoting $\eta_t^i$ as the weight that bank $i$ gives to the liquidity or the interest rate in the fitness function, such quantity becomes a function of the recent performance of the agent in terms of attractiveness. Namely, if $\mu_t^i-\mu_{t-1}^i \geq 0$, the agent $i$ intensifies the strategy she is already pursuing, then
\begin{equation}
    \eta_{t+1}^i= \begin{cases}\eta_t^i+a & \text { if } \eta_t^i \geq 0.5 \\ \eta_t^i-a & \text { if } \eta_t^i<0.5\end{cases}
\end{equation}
On the other hand, if $\mu_t^i-\mu_{t-1}^i<0$, the bank $i$ weakens the strategy she is pursuing, intensifying the opposite one
\begin{equation}
    \eta_{t+1}^i= \begin{cases}\eta_t^i-a & \text { if } \eta_t^i \geq 0.5 \\ \eta_t^i+a & \text { if } \eta_t^i<0.5\end{cases},
\end{equation}
where $a$ is a scalar parameter set equal to 0.025, defining the step towards liquidity or interest rate. We run 200 independent model simulations of length T = 1000 periods to obtain the following results. All the other agents' initialization parameters, except for the
variation studied here, overlap with those presented in Sec 3 with a percentage of followers $\kappa=1$. At $t=0$, all agents start with $\eta_{t=0}^i \sim \mathcal{U}_{[0,1]}$.

\begin{figure}
\centering
\begin{subfigure}{.5\textwidth}
  \centering
  \includegraphics[width=\linewidth]{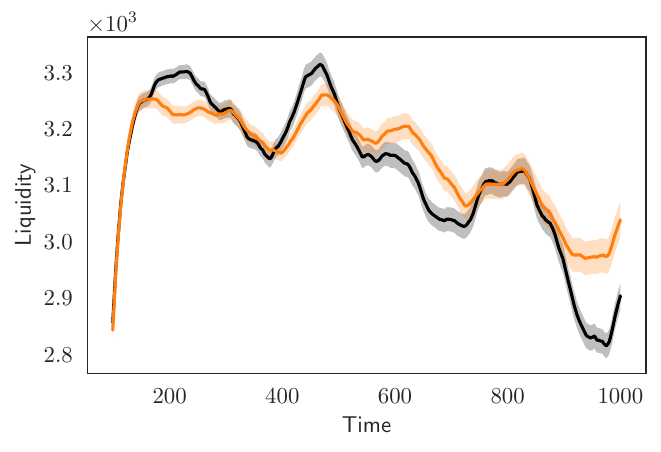}
\end{subfigure}%
\begin{subfigure}{.5\textwidth}
  \centering
  \includegraphics[width=\linewidth]{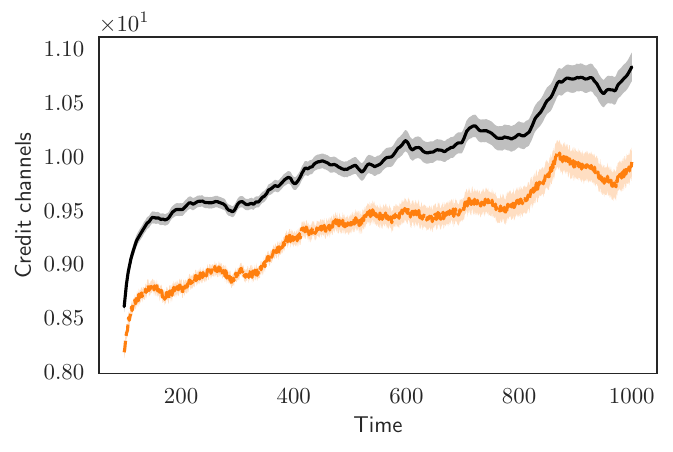}
\end{subfigure}
 \caption{Liquidity of the system (left panel) and the number of credit channels (right panel).
Black solid and orange dashed lines refer to the best-performing reinforcement learning
optimal strategy and the decentralized strategy, respectively. The curves reproduce the
mean and the standard deviation over 200 simulations of the system and a rolling
window of 100 timesteps.}
  \label{Fig:macro_liquidity_leverage_dec}
\end{figure}%
\FloatBarrier

\begin{figure}
\centering
\begin{subfigure}{.33\textwidth}
  \centering
  \includegraphics[width=\linewidth]{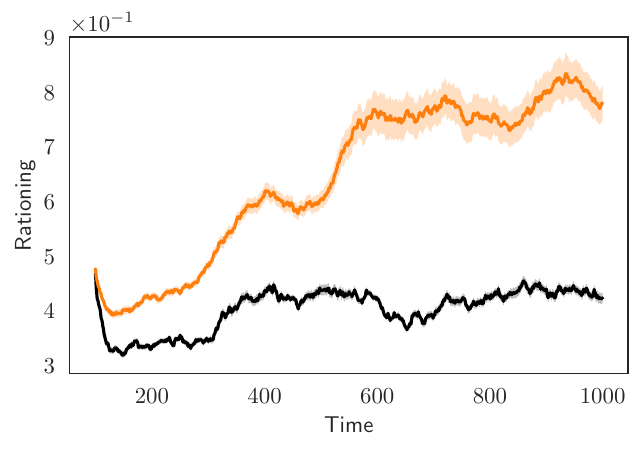}
\end{subfigure}%
\begin{subfigure}{.33\textwidth}
  \centering
  \includegraphics[width=\linewidth]{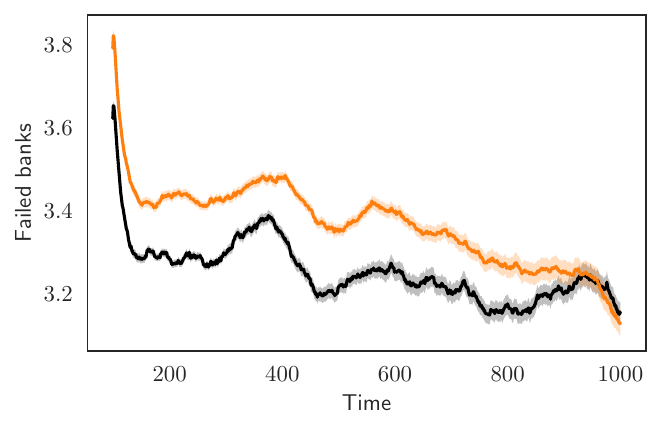}
\end{subfigure}
\centering
\begin{subfigure}{.33\textwidth}
  \centering
  \includegraphics[width=\linewidth]{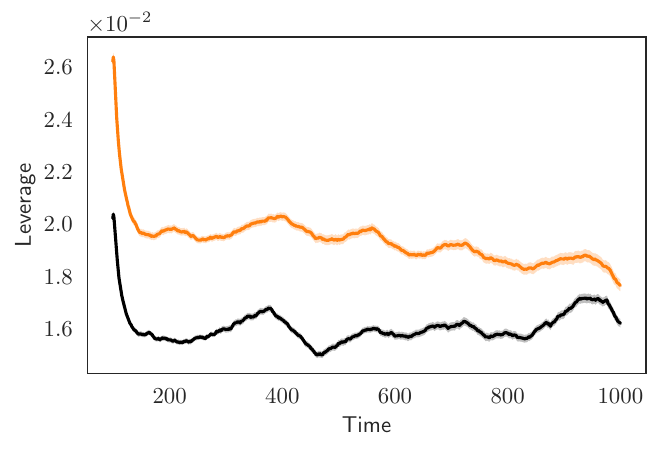}
\end{subfigure}%
 \caption{Rationing of the system (left panel), number of failed banks (middle panel), and Leverage of the system (right panel). Black solid and orange dashed lines refer to the best-performing reinforcement learning
optimal strategy and the decentralized strategy, respectively. The curves reproduce the
mean and the standard deviation over 200 simulations of the system and a rolling
window of 100 timesteps.}
  \label{Fig:macro_rationing_failed_dec}
\end{figure}
\begin{figure}
 \centering\includegraphics{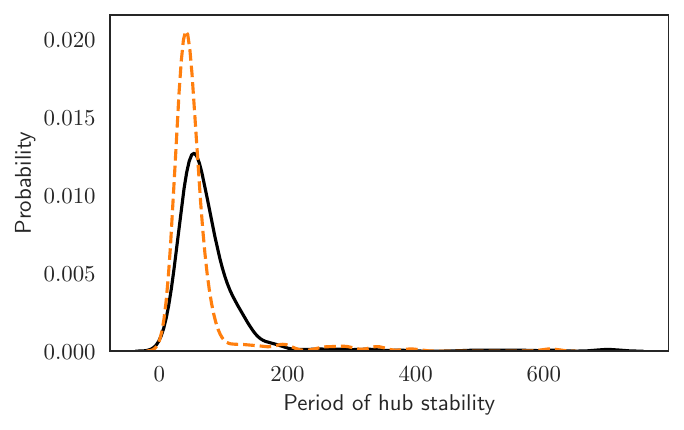}
 \vspace*{-5mm}
 \caption{Density distributions over $200$ Monte Carlo simulations of the maximum period of hub stability in which the strategy does not change. The black solid and orange dashed lines show the reinforcement learning and decentralized strategies, respectively. The black solid density is obtained by summing the two densities presented in Figure\ref{Fig:guru_dist}}
 \label{Fig:eta_hub_stability_dec}
\end{figure}

%
\FloatBarrier
As the reader can easily grasp from Fig. \ref{Fig:macro_liquidity_leverage_dec} and \ref{Fig:macro_rationing_failed_dec}, where the systemic dynamics presented in Sec. \ref{Subsec:macro} are reproduced, once again, the RL-based strategy (black solid line) generates more desirable systemic patterns than the new decentralized strategy (orange dashed line). Specifically, when the regulator adopts an $\eta$ evolving through reinforcement learning, the system absorbs shocks better than in the decentralized case, as shown by the higher number of credit channels, the lower leverage and lower rationing, and the number of failures associated with the centralized $\eta$. Only the system's liquidity trend has apparently unclear dynamics, as shown in the left panel of Fig.\ref{Fig:macro_liquidity_leverage_dec}.
\begin{figure}
 \centering\includegraphics{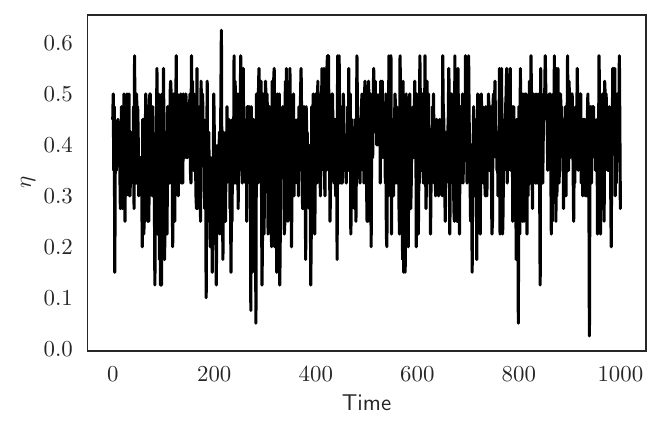}
 \vspace*{-5mm}
 \caption{Time series of the decentralized $\eta$ over the simulation. Data are obtained through 200 Monte Carlo simulations of the system.}
 \label{Fig:eta_dynamics_dec}
\end{figure}
 \FloatBarrier
To understand the reasons underlying the better systemic results obtained with a centralized versus decentralized $\eta$, we need to focus on the evolution of this variable in the two scenarios. In this regard, the top panel of Fig. \ref{fig:cumulo} and \ref{Fig:eta_dynamics_dec} show the evolution of centralized and decentralized $\eta$, respectively. The centralized strategy, i.e., the
one obtained with the reinforcement learning algorithm selects values of $\eta$ that direct the system towards the best between the two pure strategies. Given the underlying systemic conditions, such a strategy is the most advantageous by the reinforcement learning algorithm. On the contrary, the decentralized strategy shows an erratic trend in Fig. \ref{Fig:eta_dynamics_dec}. It emerges that when financial institutions choose the parameter considering their individual fitness, the system never achieves coordination in the choices. The decentralized $\eta$ dynamics also show an oscillating behavior on average between the mixed strategy ($\eta=0.5$) and the low interest rate one, with an average value of about 0.35, with a standard deviation of 0.09. The minimum and the maximum are 0 and  0.6, respectively. The lack of coordination and the approaching of the low interest rate strategy have important systemic consequences. On the one hand, the erratic nature of the decentralized strategy is not beneficial for the stability of the credit network, as confirmed in Fig.\ref{Fig:eta_hub_stability_dec} where the distributions, over 200 simulations, of the maximum period of hub stability for the reinforcement $\eta$ (black solid line) and the decentralized one (orange dashed line) are displayed. As explained in the baseline model (see Sec. \ref{Subsec:systemicimpact}), lower hub longevity indicates lower network centrality \footnote{The network centrality measure calculated over 200 Monte-Carlo simulations reaches peaks of 0.81 in the centralized case and of 0.26 in the decentralized one.}. that is associated with worse systemic performance. On the other hand, the fact that the decentralized rule comes close to the low interest rate strategy even further moves the system away from stability. In this circumstance, in fact, lenders, much smaller than borrowers, are overwhelmed in the event of their
clients' bankruptcy. Moreover, the exclusion from the exchanges of the largest institutions leaves a consistent level of unallocated liquidity in the system, which explains the apparently high liquidity in the decentralized framework shown in the red dashed line of the left panel of Fig. \ref{Fig:macro_liquidity_leverage_dec}.
 \FloatBarrier

\section{Concluding remarks}\label{Sec:fine}
This work shows the effects of a policy recommendation obtained through a reinforcement learning mechanism in an artificial interbank market. Specifically, we assume that the financial institutions receive a signal from the regulator regarding the best strategy to adopt for the creation of their lending agreements. Depending on the underlying economic conditions, the signal directs the system towards providing a high liquidity supply or a low interest rate. Using a reinforcement learning approach to provide this public signal has proven effective since the method exploits the available information and redirects the system towards an efficient flow of liquidity compared to other different static and dynamic behavioral tactics. Moreover, through the SHAP framework, which dissects the contribution of each piece of information to the recommended policy, we have been able to interpret the primary input that drives the policy choice. We have acknowledged that the occurrence of one circumstance (liquidity vs. interest rate) generates significant consequences affecting the agents' performances and the topology and resiliency of the interbank network. Specifically, when the signal directs the system toward an abundant liquidity provision, the interbank network, composed of a few populated communities, is more centralized and dense towards hub banks than in the low interest rate scenario. This network architecture accompanies better individual performances and higher system resilience in the face of exogenous shocks. Our results have shown that the better general conditions underlying this signal are due to the homogeneity between lenders and borrowers, which generates a uniform risk exposure among counterparties that favor the system's resiliency.

Leaving aside the results of the comparison between the two signals, we have analyzed the general effect of the policy recommendation implemented via the reinforcement learning procedure in the second part of the paper. Our results have shown how systemic risk is mitigated by such a tool and how this outperforms other alternative behavioral strategies. 

It is worth noting that the novelty of this work is introducing a reinforcement learning framework on top of an agent-based model by directing banks' actions towards strategies that promote systemic stability. While improving the model's realism would have captured important aspects related to stability and propagation of systemic risk, it would have compromised the explainability of the algorithm's results. Hence, given the additional layer of complexity brought by reinforcement learning, we focus on the results obtained from a simplified model whose outcomes are verifiable and controlled. Further research can expand the model to include additional agents, such as households and more sophisticated central bank.

\section*{Acknowledgments}
This work was partially supported by project SERICS (PE00000014) under the MUR National Recovery and Resilience Plan
funded by the European Union - NextGenerationEU. It was also supported by the Spanish Ministerio de Ciencia, Innovacion y Universidades (grant RTI2018-096927-B-100). DT acknowledges GNFM-Indam for financial support.
%%%%%%%%%%%%%%%%%%%%%%%%%%%%%%%%%%%%%%%%%%%
% to choose bibliography https://it.overleaf.com/learn/latex/Bibtex_bibliography_styles
\bibliographystyle{dcu}
\bibliography{ms}
%%%%%%%%%%%%%%%%%%%%%%%%%%%%%%%%%%%%%%%%%%%
\newpage
\appendix
\section{A sensitivity analysis on model parameters}\label{App:A}
In this appendix, we investigate the performances of the learning algorithm by varying some key parameters. The first investigated parameter, $\beta$, governs the network topology (see \cite{grilli2014network}, for a mathematical explanation). As the intensity of choice increases, the interbank architecture ranges from a random configuration to a star one. The effect of the network topology on the interbank system is studied by changing $\beta$ from 0 to 40 with steps of 2. The second parameter we consider is fire sale price $\rho$. An increase in $\rho$ impacts both lenders and borrowers.
On the one hand, it compensates the losses that lenders incur due to the failure of their clients (see Eq. \ref{Eq:profit}). On the other hand, a higher fire-sale increases the likelihood that the borrower, rationed in the interbank market, can face deposit repayments. Here we vary the fire-sale price, $\rho$, from 0.1 to 0.5 with steps of 0.1. Thirdly, we modify the skewness of the distribution of the random shock affecting the bank deposit at the beginning of each period.
%It is important to analyze the robustness of the learning results obtained in section \ref{Subsec:rl} when we let some system parameters vary. We identify three relevant parameters to test the effectiveness of the learning approach under different system scenarios. The first is the $\beta$ in Eq. \ref{Eq:probattachment} that reflects the intensity of breaking the preferential lending agreements through banks. The second is the price of fire-sale $\rho$, which the bank can use as a last resort to collect the needed money by dismantling their long-term asset at a discount. The third refers to the magnitude of the random shock that modifies the bank deposit at the beginning of each period. 
Recalling the equation for the deposit movements as $D_t^i=D_{t-1}^i (\mu + \omega U(0,1))$, we remark that it allows us to reproduce bearish and bullish market periods. The uniformly distributed noise component can be shifted towards more negative or positive shocks at convenience to represent different market situations. Having fixed $\mu=0.7$ in our simulations, we let $\omega$ vary from 0.52 to 0.6 with steps of 0.02, corresponding to a highly negatively skewed and perfectly symmetrical shock distribution. 

The role of $\mu$ and $\omega$ is critical to regulating the magnitude of the aggregated shock that affects the interbank system. Precisely, $\mu$ and $\omega$ determine the probability of the sign of the deposit's shock. When  $\mu=0.7$ and $\omega=0.6$, the likelihood of a negative shock is equal to that of a positive one. This parameter configuration corresponds to an interbank market meeting the
conditions of accounting consistencies, where on average, the other half of the market participants recover what is eroded by the adverse market condition. In this circumstance, the total number of assets for each bank matches the total number of liabilities, hence the aggregated balance sheet of the system sum to 0 at the beginning of each  trading day. We refer to accounting consistency rather than stock-flow consistency, because the latter is more appropriate to multi-sector macroeconomic models such as those proposed by \cite{caiani2014innovation} and \cite{caiani2016agent}). In our baseline model, we used $\mu=0.7$ and $\omega=0.55$ to favor more adverse shocks and, therefore, to have more interbank market activity to cover such needs.

The last part of this appendix is dedicated to investigating the effects of a change in the reserve ratio, $\hat{r}$, previously set at 2\%. This analysis has a twofold value. On the one hand, it is a further experiment on the robustness of the model by changing the parameters space. On the other, it corresponds to a conventional monetary policy. %the comparison between $\eta$  evolving via the reinforcement learning mechanism and the two fixed strategies, i.e., a constant $\eta$ equal to 0 and 1, respectively.

In all these experiments, we run our model 100 times for different values of the initial seed generating the pseudo-random numbers over a time span of T = 1000 periods. Moreover, all the agents' initialization parameters, except for the variations studied here, coincide with those presented in Sec. \ref{Sec:simulation}.   

\begin{figure}
\centering
\begin{subfigure}{.33\textwidth}
  \centering
  \includegraphics[width=\linewidth]{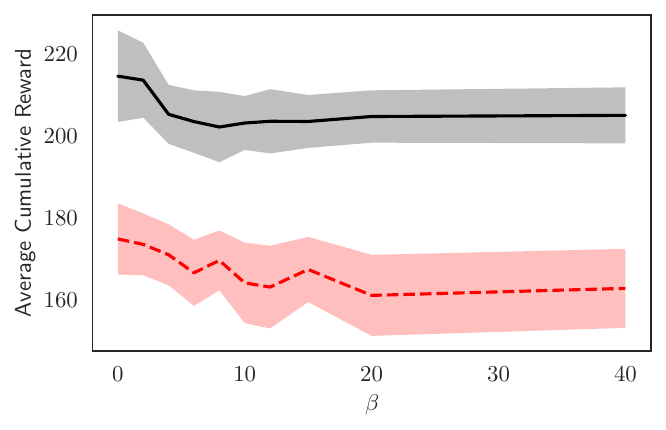}
\end{subfigure}%
\begin{subfigure}{.33\textwidth}
  \centering
  \includegraphics[width=\linewidth]{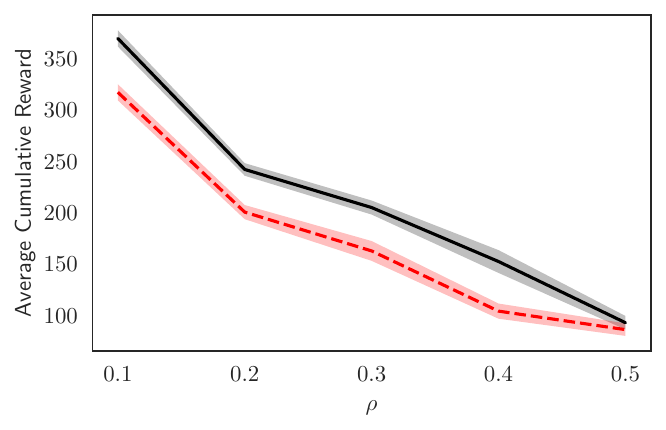}
\end{subfigure}
\begin{subfigure}{.33\textwidth}
  \centering
  \includegraphics[width=\linewidth]{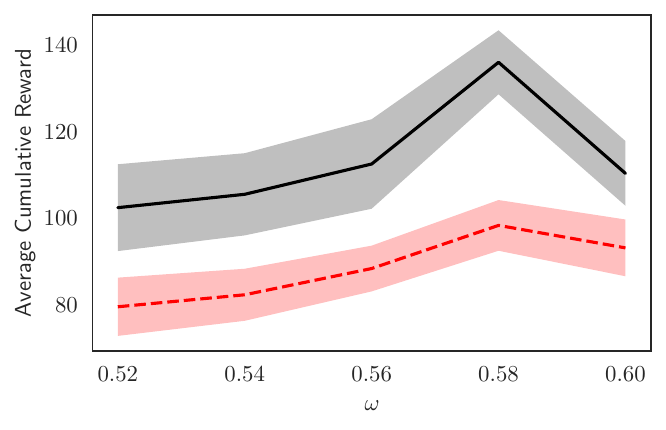}
\end{subfigure}
\caption{Average cumulative fitness of the system as a function of changes in $\beta$, $\rho$, and $\omega$ in the first, second, and third panels, respectively. The reinforcement learning algorithm is in solid black, while the random strategy is in dashed red.}
\label{Fig:reward_sensitivity}
\end{figure}

Let us begin the analysis by focusing on the three-parameter variations' implications on the model's results. Each parameter variation represents a different configuration of the banking system, which is used to test the different strategies over 100 simulations. The cumulative reward of these simulations is then averaged to obtain the mean values and the respective confidence interval for the reinforcement learning strategy and the random strategy. Fig. \ref{Fig:reward_sensitivity} shows the average cumulative reward over the 100 simulations as a function of a single parameter variation. We notice that the performance of the reinforcement learning algorithm solved with the PPO procedure is still superior with respect to the random strategy for all three sensitivity cases presented. Therefore, we can conclude that the effect analysis in the main paper still holds if one modifies some characteristics of the underlying financial system.

\FloatBarrier

\begin{figure}
\centering
%%%%%%%%%%%%%%%%%%%%%%%%%%%%%%%%%%%%%%%%%%%%%%%%%%%%%%%%%%%%%%%%%%%%%%%%%% 
\begin{subfigure}{.33\textwidth}
  \centering
  \includegraphics[width=\linewidth]{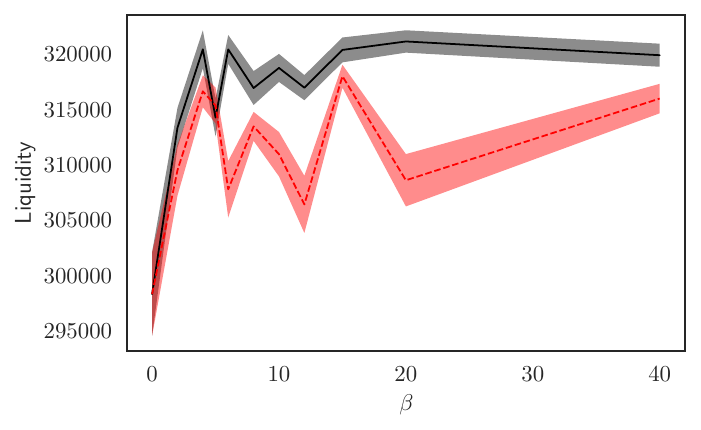}
\end{subfigure}%
\begin{subfigure}{.33\textwidth}
  \centering
  \includegraphics[width=\linewidth]{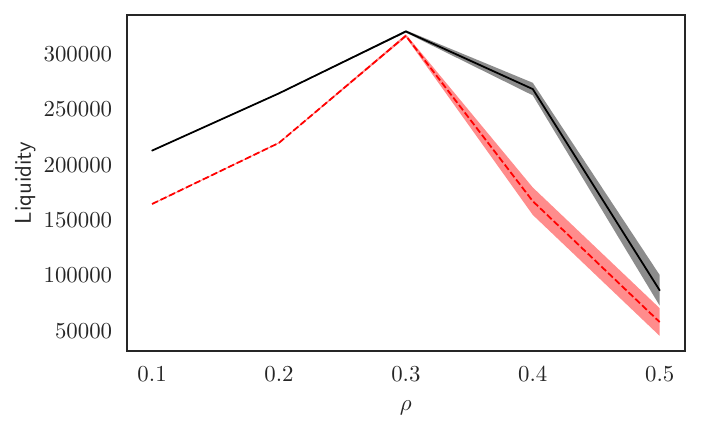}
\end{subfigure}
\begin{subfigure}{.33\textwidth}
  \centering
  \includegraphics[width=\linewidth]{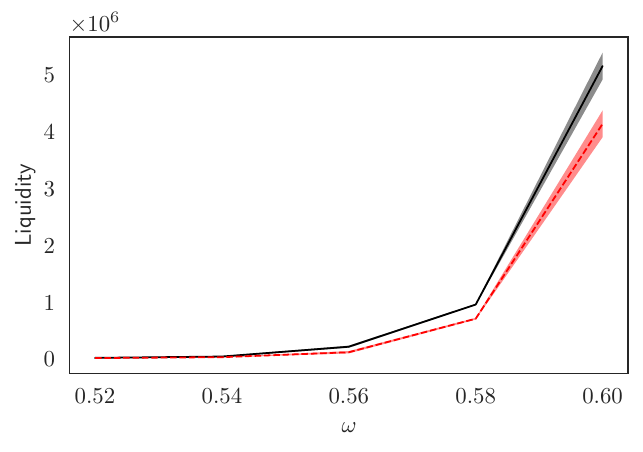}
\end{subfigure}
%%%%%%%%%%%%%%%%%%%%%%%%%%%%%%%%%%%%%%%%%%%%%%%%%%%%%%%%%%%%%%%%%%%%%%%%%%
\begin{subfigure}{.33\textwidth}
  \centering
  \includegraphics[width=\linewidth]{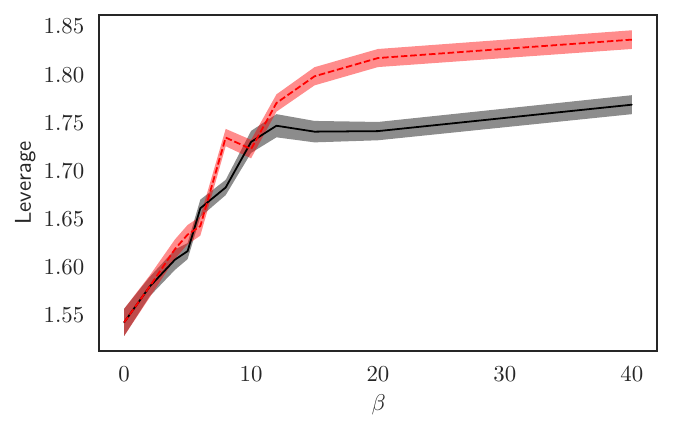}
\end{subfigure}%
\begin{subfigure}{.33\textwidth}
  \centering
  \includegraphics[width=\linewidth]{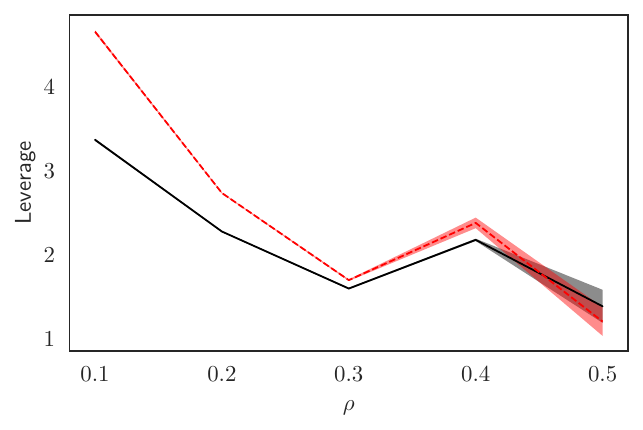}
\end{subfigure}
\begin{subfigure}{.33\textwidth}
  \centering
  \includegraphics[width=\linewidth]{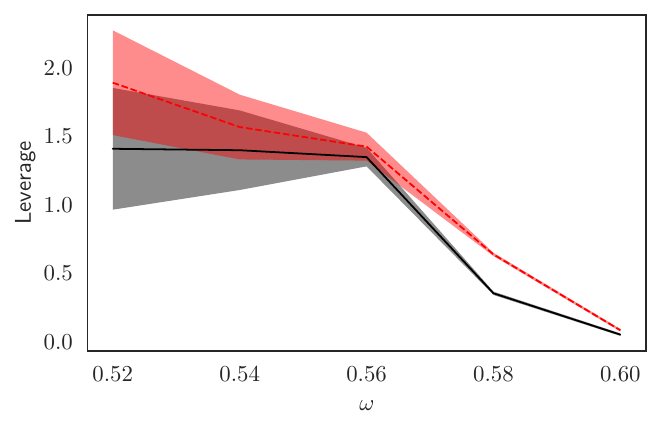}
\end{subfigure}
%%%%%%%%%%%%%%%%%%%%%%%%%%%%%%%%%%%%%%%%%%%%%%%%%%%%%%%%%%%%%%%%%%%%%%%%%%
% \begin{subfigure}{.33\textwidth}
%   \centering
%   \includegraphics[width=\linewidth]{beta_baddebt_sensitivity_3_2sec.pdf}
%   \label{fig:sub1}
% \end{subfigure}%
% \begin{subfigure}{.33\textwidth}
%   \centering
%   \includegraphics[width=\linewidth]{fire_sale_price_baddebt_sensitivity_3_2sec.pdf}
%   \label{fig:sub2}
% \end{subfigure}
% \begin{subfigure}{.33\textwidth}
%   \centering
%   \includegraphics[width=\linewidth]{shock_volatility_baddebt_sensitivity_3_2sec.pdf}
%   \label{fig:sub2}
% \end{subfigure}
%%%%%%%%%%%%%%%%%%%%%%%%%%%%%%%%%%%%%%%%%%%%%%%%%%%%%%%%%%%%%%%%%%%%%%%%%%
\begin{subfigure}{.33\textwidth}
  \centering
  \includegraphics[width=\linewidth]{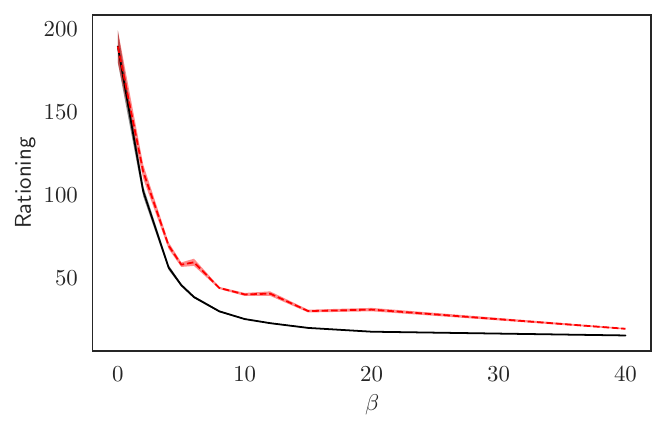}
\end{subfigure}%
\begin{subfigure}{.33\textwidth}
  \centering
  \includegraphics[width=\linewidth]{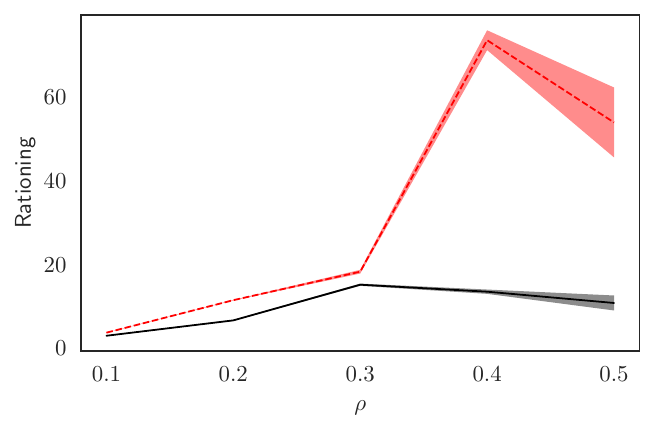}
\end{subfigure}
\begin{subfigure}{.33\textwidth}
  \centering
  \includegraphics[width=\linewidth]{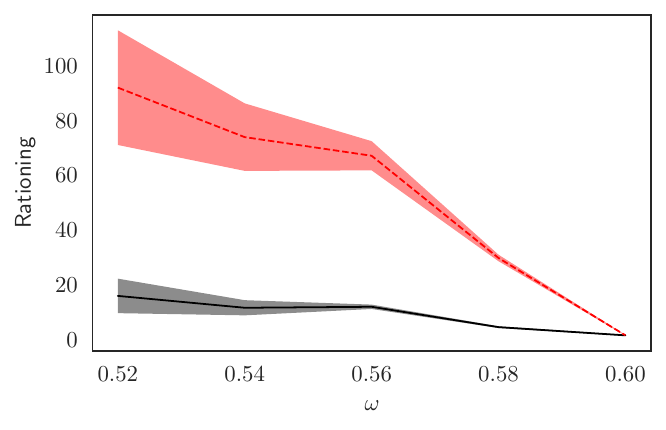}
\end{subfigure}
%%%%%%%%%%%%%%%%%%%%%%%%%%%%%%%%%%%%%%%%%%%%%%%%%%%%%%%%%%%%%%%%%%%%%%%%%%
\begin{subfigure}{.33\textwidth}
  \centering
  \includegraphics[width=\linewidth]{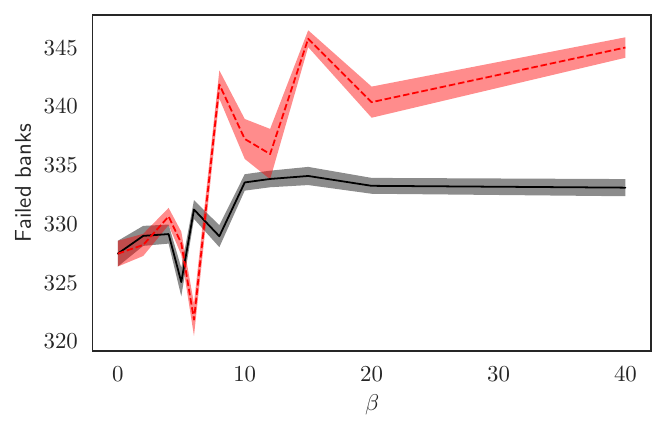}
\end{subfigure}%
\begin{subfigure}{.33\textwidth}
  \centering
  \includegraphics[width=\linewidth]{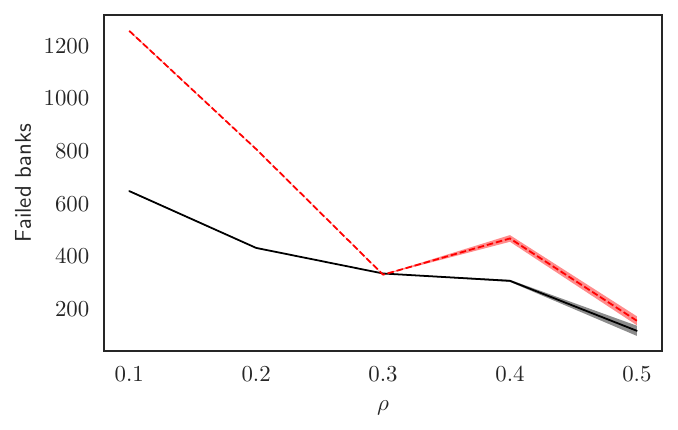}
\end{subfigure}
\begin{subfigure}{.33\textwidth}
  \centering
  \includegraphics[width=\linewidth]{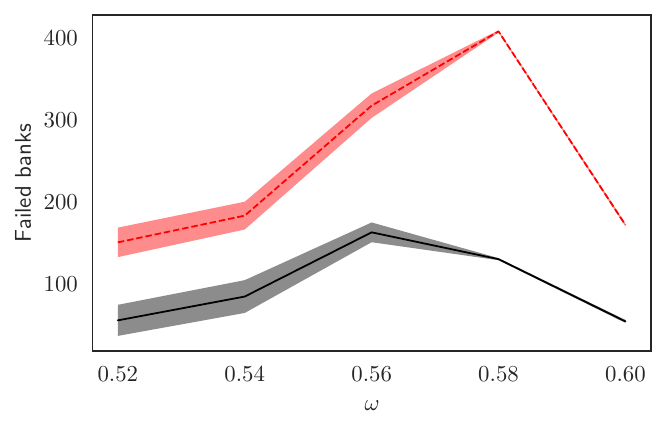}
\end{subfigure}
%%%%%%%%%%%%%%%%%%%%%%%%%%%%%%%%%%%%%%%%%%%%%%%%%%%%%%%%%%%%%%%%%%%%%%%%%%
\begin{subfigure}{.33\textwidth}
  \centering
  \includegraphics[width=\linewidth]{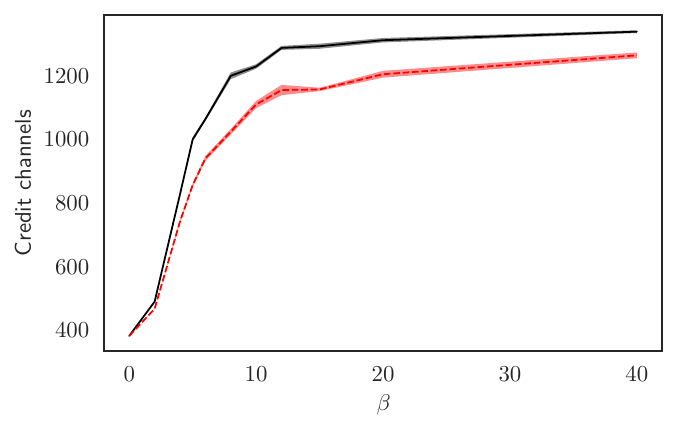}
\end{subfigure}%
\begin{subfigure}{.33\textwidth}
  \centering
  \includegraphics[width=\linewidth]{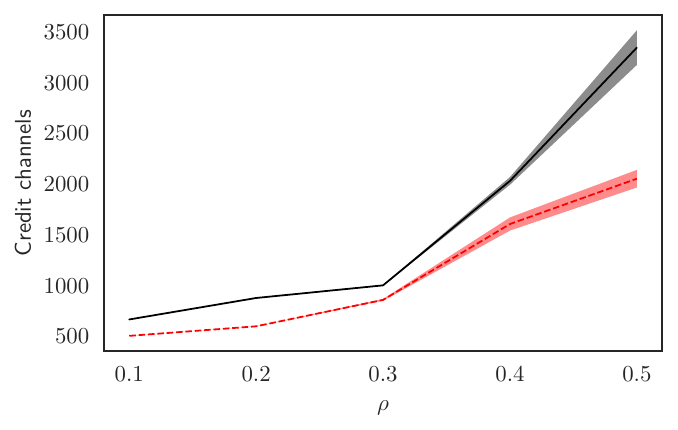}
\end{subfigure}
\begin{subfigure}{.33\textwidth}
  \centering
  \includegraphics[width=\linewidth]{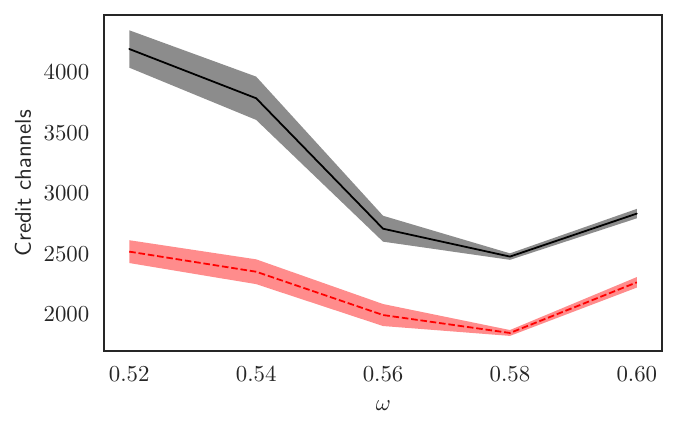}
\end{subfigure}
%%%%%%%%%%%%%%%%%%%%%%%%%%%%%%%%%%%%%%%%%%%%%%%%%%%%%%%%%%%%%%%%%%%%%%%%%%
\vspace*{-5mm}
\caption{Sensitivity analysis on system variables in the face of changes in $\beta$, $\rho$, and $\omega$, in the first, second and third columns, respectively. The reinforcement learning algorithm is in solid black, while the random strategy is in dashed red.}
\label{Fig:macro_sensitivity}
\end{figure}

In Fig. \ref{Fig:macro_sensitivity} we show the sensitivity of the average values, over all the 100 simulations and a rolling window of 100 timesteps, of relevant quantities at the systemic level with respect to the three parameters described above\footnote{We refer the reader to Sec. \ref{Subsec:macro} for a detailed explanation on the implementation of Fig. \ref{Fig:macro_sensitivity}}. Before going into the details concerning the systemic impact of the single parameters, we can observe that the reinforcement learning strategy (solid black in Fig.\ref{Fig:macro_sensitivity}) consistently outperforms the random strategy (dashed red line in Fig.\ref{Fig:macro_sensitivity}) over all parameters and variables considered\footnote{To appreciate the statistical significance of the reinforcement learning strategy with respect to the random strategy, we performed a series of T-tests for each variable in the figures presented. The results show a statistically significant difference between each pair of curves at least the 5\% level. We omitted here the table, including the p-values that are available under requests, as well as the results of the sensitivity analysis that we performed on the parameters $\hat{d}$, $\chi$, $\phi$ and $\xi$.}. The system generated with the reinforcement learning algorithm produces, on the one hand, higher liquidity and more credit channels and, on the other hand, lower rationing, bankruptcies, and leverage than the one with the random algorithm.\\ Let us now analyze how variations in each parameter affect the system's stability. In the first column of Fig. \ref{Fig:macro_sensitivity}, we show the effects that the intensity of choice, $\beta$, has on the systemic variables. When $\beta$ increases from 0 to 40, the liquidity and the credit channels increase to $\beta=10$ and stabilize. This pattern occurs in both scenarios (i.e., with optimal and random strategies). The underlying reason for this dynamic is as follows: a $\beta$ value greater than or equal to 10 generates a stable topology in the interbank network, which makes the investigated values insensitive to further changes in the parameter. Similar to the trend of the previous variables are the leverage dynamics, which increase with $\beta$ but at a decreasing rate,  which is confirmed for both the adopted strategies.
Indeed, the more liquidity is available in the system, the more exchange of loans between banks happens. Finally, an increasing $\beta$ causes the amount of rationing of the system to decrease in both the considered scenarios, while the failures of the agent happen to be stable over the period under the optimal strategy or increase under the random scenario.\\ In the second column of Fig. \ref{Fig:macro_sensitivity}, we focus on the effects produced by a variation in the fire-sale price. An increase of $\rho$ protects both lenders and borrowers from losses, and it is beneficial when looking at the liquidity up to $\rho=0.3$. From that level, borrowers do not enter the interbank market frequently because they can cover their needs by selling their long-term assets at a satisfactory price. This is also reflected in the amount of rationing and failures that decrease when $\rho$ is above 0.3. The leverage immediately decreases with $\rho$ because the increase in the system's liquidity is more than compensated by the increase in equity since lenders are usually repaid by borrowers and do not lose parts of their equity. The dynamics produced by the fire-sale price variation are valid when observing the system with the optimal signal and the one with the random signal. Finally, in the last column of the figure, the impact of the deposit's motion is investigated. The increase of the $\omega$ parameter causes an increase in liquidity since the shocks become gradually less and less harmful. This also explains the decrease in the leverage and the rationing because banks are less negatively impacted by the deposit shock and, consequently, need to gather less money from the market. For the same reason, the amount of credit channels decreases with a more symmetric shock distribution. In contrast, the failures are substantially stable, except for a higher variability when $\omega$ describes a highly asymmetric shock. Also, for this last parameter, the system dynamics produced with the optimal signal follow the same trend as those obtained with the random signal.

\FloatBarrier

\begin{figure}
\begin{subfigure}{.33\textwidth}
  \centering
  \includegraphics[width=\linewidth]{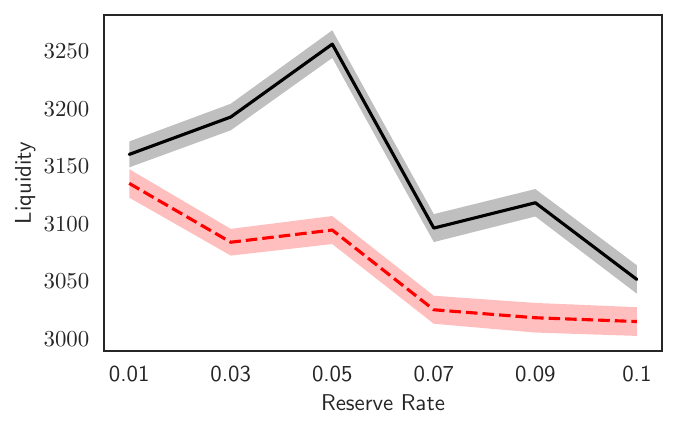}
\end{subfigure}%
\begin{subfigure}{.33\textwidth}
  \centering
  \includegraphics[width=\linewidth]{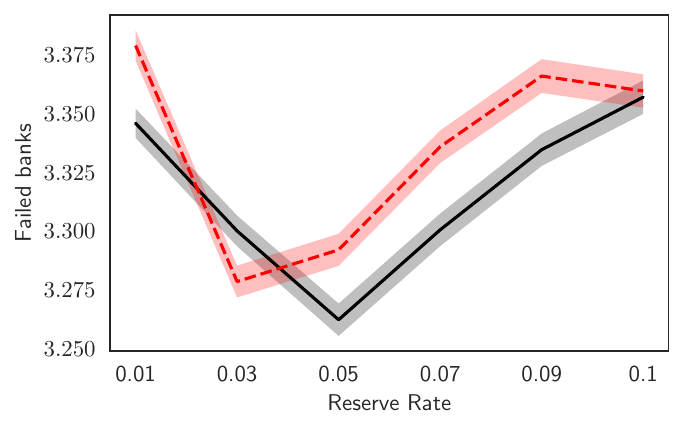}
\end{subfigure}
\begin{subfigure}{.33\textwidth}
  \centering
  \includegraphics[width=\linewidth]{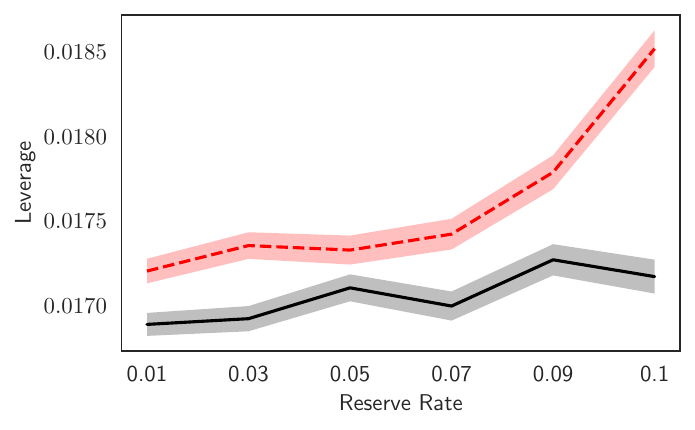}
\end{subfigure}

\caption{Sensitivity analysis on the liquidity, number of bankruptcies, and leverage in the face of changes in reserve ratio $\hat{r}$ in the first, second, and third columns, respectively. The reinforcement learning algorithm is in solid black, while the random strategy is in dashed red.}
\label{reserve_r}
\end{figure}

 In the last part of our analysis, we study how the system's resilience varies as the reserve requirement ratio varies from 1\% to 10\%. Fig\ref{reserve_r} shows the sensitivity of the average values over all the 100 simulations and a rolling window of 100 timesteps of relevant quantities at the systemic level with respect to the variation of $\hat{r}$. Before describing the effects of the contractionary monetary policy on market stability, it is worth noting that the system obtained through the optimized $\eta$ (solid black line) consistently outperforms that with the random signal (dashed red line). The former always generate higher liquidity, lower leverage, and several failures. If we now observe the systemic effects of the increase in reserve ratio in the framework with the optimized signal, we can see an inverted U-shaped trend in liquidity. For $\hat{r}$-values between 1\% and 5\% , liquidity increases, showing that a non-excessively high reserve ratio promotes interbank stability by decreasing the number of failures. However, when the central bank imposes a reserve ratio above 5\%, the contractionary effect of the policy takes over. The system becomes less liquid, and this causes a spike in failures as banks can no longer cope with their adverse deposit shocks. Finally, the behavior of the leverage, always in the context of the optimal signal, is timidly monotonically increasing with $\hat{r}$ (see black line in the right-hand panel of Fig.\ref{reserve_r}). For values of $\hat{r}$ up to 5\%, the leverage increases due to the rise in the granting of a loan. Above this threshold, the increase in leverage is mainly caused by the higher number of bankruptcies, which negatively impacts the net worth of financial institutions.

%%%%%%%%%%%%%%%%%%%%%%%%%%%%%%%%%%%%%%%%%%%%%%%%%%%%%%%%%%%%%%%%%%%%%%%%%%%%%%%%
\section{Algorithms and Hyperparameters}\label{App:B}
The PPO algorithm is easier to implement than a trust-region method \citep{schulman2015trust} and easier to tune with respect to of Deep-Q network (DQN) \cite{mnih2015human} or its continuous counterpart \citep{lillicrap2015continuous}. Our implementation of PPO follows \cite{andrychowicz2020matters}, which performs an extensive empirical study of the effect of implementation and parameter choices on PPO performances. Even if we use the algorithm in a different context than their testbed, we follow the direction of their results in order to tune our hyperparameters. 
% Our implementation is available through a GitHub repository (\alessio{Inserire qui la repo the renderò Pubblica Quando avremo il working paper finito}).

As described in the main, we implement PPO in an actor-critic setting without shared architectures. When used to parametrize discrete strategies, policy gradient methods like PPO output a normalized set of logits to get the corresponding probabilities. Then, a greedy strategy selects the action which obtains the maximum probability. The entropy bonus guarantees exploration during training in the objective function.

The on-policy feature of PPO makes the training process episodic so that experience is collected by interacting with the environment and then discarded immediately once the strategy has been updated. In principle, on-policy learning appears a more obvious learning setup, even if it comes with some caveats. It makes the training less sample efficient and computationally expensive since a new sequence of experiences must be collected after each update step. In this process, the advantage function is computed before the optimization steps, when the discounted sum of returns over the episode can be computed. In order to increase the training efficiency, after one sweep through the collected samples, we compute the advantage estimator again and perform another sweep through the same experience. This trick reduces the computational expense of recollecting experiences and increases the sample efficiency of the training process. Usually, we do at most three sweeps (epochs) over a set of collected experiences before moving on and collecting a new set.

The gradient descent optimizer is Adam (\cite{kingma2014adam}), which performs a batch update of size 100 with a learning rate of 0.005. Since the data are not all available in a reinforcement learning setting at the beginning of the training, we can not normalize our input variables as usual in the preprocessing step of a supervised learning context. Hence, we add a Batch Normalization layer (\cite{ioffe2015batch}) before the first hidden layer to normalize the inputs batch by batch and obtain the same effect. 

Maximizing the objective function that returns the gradient in Eq. \ref{Eq:pgadvantage} is unstable since updates are not bounded and can move the strategy too far from the local optimum. Similarly to TRPO (\cite{schulman2015trust}), PPO optimizes an alternative objective to mitigate the instability
\begin{equation}\label{Eq:CLIP}
J^{\text{CLIP}}(\theta,\psi) = \mathbb{E}_{\pi_{\theta}} \left[ \min \left( r(\theta) \hat{\mathbb{A}}\left(s,a;\psi\right), \operatorname{clip}\left(r(\theta), 1-\epsilon, 1+\epsilon\right) \hat{\mathbb{A}}\left(s,a;\psi\right) \right) \right]
\end{equation}
where $r(\theta) = \frac{\pi\left(A_{t} \mid S_{t}; \theta\right)}{\pi\left(A_{t} \mid S_{t}; \theta_{\text {old}}\right)}$ is a ratio indicating the relative probability of an action under the current strategy with respect to the old one. Instead of introducing a hard constraint as in TRPO, the ratio is bounded according to a tolerance level $\epsilon$ to limit the magnitude of the updates. The combined objective function in Eq. \ref{Eq:ppoobj} can be easily optimized by the PyTorch's automatic differentiation engine, which quickly computes the gradients with respect to the two sets of parameters $\theta$ and $\psi$. The implemented advantage estimator depends on the parameterized value function $V_{\psi}$ and is a truncated version of the one introduced by (\cite{mnih2016asynchronous}) for a rollout trajectory (episode) of length $T$:
\begin{equation}
\hat{\mathbb{A}}_{t}=\delta_{t}+(\gamma \tau) \delta_{t+1}+\cdots+\cdots+(\gamma \tau)^{T-t+1} \delta_{T-1} 
\end{equation}
where $\delta_{t}=r_{t}+\gamma V_{\psi}\left(s_{t+1}\right)-V_{\psi}\left(s_{t}\right)$, $\gamma$ is a discount rate with the same role of $\rho$ in DQN and $\tau$ is the exponential weight discount which controls the bias-variance trade-off in the advantage estimation. The generalized advantage estimator (GAE) uses a discounted sum of temporal difference residuals.

\end{document}